\documentclass[twocolumn,aps,prl,10pt,superscriptaddress,longbibliography,floatfix,citeautoscript]{revtex4-2}

\usepackage{amsmath}
\usepackage{amssymb}
\usepackage{bm}
\usepackage{dcolumn}
\usepackage{glossaries}
\usepackage{graphicx}
\usepackage{multirow}
\usepackage{fancyvrb}
\usepackage[linesnumbered,ruled]{algorithm2e}
\usepackage{siunitx}
\usepackage[english]{babel} 
\usepackage[usenames,dvipsnames,svgnames]{xcolor}
\usepackage{hyperref}
\hypersetup{
    pdfnewwindow=true,      
    colorlinks=true,        
    linkcolor=Blue,         
    citecolor=Blue,         
    filecolor=Blue,         
    urlcolor=Blue           
}

\usepackage{listings}	
\newacronym{bcc}{BCC}{body-centered cubic}
\newacronym{dft}{DFT}{density-functional theory}
\newacronym{dp}{DP}{deep potential}
\newacronym{eam}{EAM}{embedded-atom method}
\newacronym{fcc}{FCC}{face-centered cubic}
\newacronym{hcp}{HCP}{hexagonal close packed}
\newacronym{mae}{MAE}{mean absolute error}
\newacronym{mc}{MC}{Monte Carlo}
\newacronym{mcmd}{MCMD}{hybrid Monte Carlo and molecular dynamics}
\newacronym{md}{MD}{molecular dynamics}
\newacronym{ml}{ML}{machine learning}
\newacronym{mlp}{MLP}{machine-learned potential}
\newacronym{mpea}{MPEA}{multi-principal-element alloy}
\newacronym{nep}{NEP}{neuroevolution potential}
\newacronym{nn}{NN}{neural network}
\newacronym{pc}{PC}{principal component}
\newacronym{rmse}{RMSE}{root-mean-square error}
\newacronym{snes}{SNES}{separable natural evolution strategy}
\newacronym{unep1}{UNEP-v1}{version 1 of unified NEP}
\newacronym{zbl}{ZBL}{Ziegler-Biersack-Littmark}
\newacronym{atomsk}{ATOMSK}{The Swiss-army knife of atomic simulations}
\newacronym{ovito}{OVITO}{ Open Visualization Tool}

\DeclareSIUnit\angstrom{\text{Å}}
\DeclareSIUnit{\atom}{atom}
\DeclareSIUnit{\step}{step}
\DeclareSIUnit{\atomstepsecond}{\atom\step\per\second}
\DeclareSIUnit{\gpu}{GPU}

\newcolumntype{d}{D{.}{.}{-1}}

\makeatletter
\newcommand{\manuallabel}[2]{\def\@currentlabel{#2}\label{#1}}
\makeatother

\newcounter{note}
\newcommand{\notetitlelabel}[2]{
    \phantomsection
    \refstepcounter{note}
    \addcontentsline{toc}{subsection}{\ref{#1}. #2}
    \manuallabel{#1}{\thenote}
    \subsection*{Supplementary Note \thenote: #2}
}

\begin{document}

\title{General-purpose machine-learned potential for 16 elemental metals and their alloys}

\author{Keke Song}
\thanks{These authors contributed equally to this work.}
\affiliation{Beijing Advanced Innovation Center for Materials Genome Engineering,  University of Science and Technology Beijing, Beijing 100083, P. R. China}

\author{Rui Zhao}
\thanks{These authors contributed equally to this work.}
\affiliation{School of Materials Science and Engineering, Hunan University, Changsha 410082, China}

\author{Jiahui Liu}
\thanks{These authors contributed equally to this work.}
\affiliation{Beijing Advanced Innovation Center for Materials Genome Engineering,  University of Science and Technology Beijing, Beijing 100083, P. R. China}

\author{Yanzhou Wang}
\affiliation{MSP group, QTF Centre of Excellence, Department of Applied Physics, P.O. Box 15600, Aalto University, FI-00076 Aalto, Espoo, Finland}
\affiliation{Beijing Advanced Innovation Center for Materials Genome Engineering,  University of Science and Technology Beijing, Beijing 100083, P. R. China}

\author{Eric Lindgren}
\affiliation{
  Chalmers University of Technology,
  Department of Physics,
  41926 Gothenburg, Sweden
}

\author{Yong Wang}
\affiliation{National Laboratory of Solid State Microstructures, School of Physics and Collaborative Innovation Center of Advanced Microstructures, Nanjing University, Nanjing 210093, P. R. China}

\author{Shunda Chen}
\email{phychensd@gmail.com}
\affiliation{Department of Civil and Environmental Engineering, George Washington University,
Washington, DC 20052, USA}

\author{Ke Xu}
\affiliation{Department of Electronic Engineering and Materials Science and Technology Research Center, The Chinese University of Hong Kong, Shatin, N.T., Hong Kong SAR, 999077, P. R. China}

\author{Ting Liang}
\affiliation{Department of Electronic Engineering and Materials Science and Technology Research Center, The Chinese University of Hong Kong, Shatin, N.T., Hong Kong SAR, 999077, P. R. China}

\author{Penghua Ying}
\affiliation{Department of Physical Chemistry, School of Chemistry, Tel Aviv University, Tel Aviv, 6997801, Israel}

\author{Nan Xu}
\affiliation{Institute of Zhejiang University-Quzhou, Quzhou 324000, P. R. China}
\affiliation{College of Chemical and Biological Engineering, Zhejiang University, Hangzhou 310027, P. R. China}

\author{Zhiqiang Zhao}
\affiliation{State Key Laboratory of Mechanics and Control of Mechanical Structures, Key Laboratory for Intelligent Nano Materials and Devices of Ministry of Education, and Institute for Frontier Science, Nanjing University of Aeronautics and Astronautics, Nanjing 210016, P. R. China}

\author{Jiuyang Shi}
\affiliation{National Laboratory of Solid State Microstructures, School of Physics and Collaborative Innovation Center of Advanced Microstructures, Nanjing University, Nanjing 210093, P. R. China}

\author{Junjie Wang}
\affiliation{National Laboratory of Solid State Microstructures, School of Physics and Collaborative Innovation Center of Advanced Microstructures, Nanjing University, Nanjing 210093, P. R. China}

\author{Shuang Lyu}
\affiliation{Department of Mechanical Engineering, The University of Hong Kong, Pokfulam Road, Hong Kong SAR, P. R. China}

\author{Zezhu Zeng}
\affiliation{Department of Mechanical Engineering, The University of Hong Kong, Pokfulam Road, Hong Kong SAR, P. R. China}

\author{Shirong Liang}
\affiliation{School of Science, Harbin Institute of Technology, Shenzhen, 518055, P. R. China}

\author{Haikuan Dong}
\affiliation{College of Physical Science and Technology, Bohai University, Jinzhou 121013, P. R. China}

\author{Ligang Sun}
\affiliation{School of Science, Harbin Institute of Technology, Shenzhen, 518055, P. R. China}  

\author{Yue Chen}
\affiliation{Department of Mechanical Engineering, The University of Hong Kong, Pokfulam Road, Hong Kong SAR, P. R. China}

\author{Zhuhua Zhang}
\affiliation{State Key Laboratory of Mechanics and Control of Mechanical Structures, Key Laboratory for Intelligent Nano Materials and Devices of Ministry of Education, and Institute for Frontier Science, Nanjing University of Aeronautics and Astronautics, Nanjing 210016, P. R. China}

\author{Wanlin Guo}
\affiliation{State Key Laboratory of Mechanics and Control of Mechanical Structures, Key Laboratory for Intelligent Nano Materials and Devices of Ministry of Education, and Institute for Frontier Science, Nanjing University of Aeronautics and Astronautics, Nanjing 210016, P. R. China}

\author{Ping Qian}
\affiliation{Beijing Advanced Innovation Center for Materials Genome Engineering, University of Science and Technology Beijing, Beijing 100083, P. R. China}

\author{Jian Sun}
\email{jiansun@nju.edu.cn}
\affiliation{National Laboratory of Solid State Microstructures, School of Physics and Collaborative Innovation Center of Advanced Microstructures, Nanjing University, Nanjing 210093, P. R. China}

\author{Paul Erhart}
\email{erhart@chalmers.se}
\affiliation{
  Chalmers University of Technology,
  Department of Physics,
  41926 Gothenburg, Sweden
}

\author{Tapio Ala-Nissila}
\affiliation{MSP group, QTF Centre of Excellence, Department of Applied Physics, P.O. Box 15600, Aalto University, FI-00076 Aalto, Espoo, Finland}
\affiliation{Interdisciplinary Centre for Mathematical Modelling, Department of Mathematical Sciences, Loughborough University, Loughborough, Leicestershire LE11 3TU, UK}

\author{Yanjing Su}
\email{yjsu@ustb.edu.cn}
\affiliation{Beijing Advanced Innovation Center for Materials Genome Engineering,  University of Science and Technology Beijing, Beijing 100083, P. R. China}

\author{Zheyong Fan}
\email{brucenju@gmail.com}
\affiliation{College of Physical Science and Technology, Bohai University, Jinzhou 121013, P. R. China}

\date{\today}

\begin{abstract}
Machine-learned potentials (MLPs) have exhibited remarkable accuracy, yet the lack of general-purpose MLPs for a broad spectrum of elements and their alloys limits their applicability. Here, we present a feasible approach for constructing a unified general-purpose MLP for numerous elements, demonstrated through a model (UNEP-v1) for 16 elemental metals and their alloys.
To achieve a complete representation of the chemical space, we show, via principal component analysis and diverse test datasets, that employing one-component and two-component systems suffices. 
Our unified UNEP-v1 model exhibits superior performance across various physical properties compared to a widely used embedded-atom method potential, while maintaining remarkable efficiency. 
We demonstrate our approach's effectiveness through reproducing experimentally observed chemical order and stable phases, and large-scale simulations of plasticity and primary radiation damage in MoTaVW alloys. 
This work represents a significant leap towards a unified general-purpose MLP encompassing the periodic table, with profound implications for materials science.
\end{abstract}

\maketitle


Atomistic simulations of elemental metals and their alloys play a crucial role in understanding and engineering materials properties.
While quantum-mechanical methods such as \gls{dft} calculations can be directly used for small simulation cells and short sampling times, their feasibility quickly diminishes with increasing spatial and temporal scales.
For large-scale classical atomistic simulations, both \gls{md} and \gls{mc} simulations crucially depend on interatomic potentials.
For metallic systems in particular, \gls{eam}-type potentials \cite{daw1984prb, Finnis1984pma} have proven to be useful and been extensively applied over the past decades, especially for elemental metals and their alloys.
However, these existing classical interatomic potentials often lack the required level of accuracy for numerous applications.
This deficiency primarily stems from constrained functional forms.
Recently, a novel paradigm for developing interatomic potentials has emerged based on \gls{ml} techniques \cite{behler2016jcp, Deringer2019am, Mueller2020jcp, Noe2020arpc, Mishin2021am, Unke2021cr}.
In a \gls{mlp}, the interatomic potential is modeled using \gls{ml} methods, allowing for a significantly greater number of fitting parameters and providing versatility as compared to traditional many-body potentials.
The functional forms of these \glspl{mlp} are remarkably flexible, free from the limitations of a small number of analytical functions suggested by physical and chemical intuition or fitting to ground state properties only.
The combination of flexible functional forms and a large number of fitting parameters empowers \glspl{mlp} to achieve a level of accuracy that can be well beyond that of the traditional many-body potentials.

The basic theory behind \glspl{mlp} is rather mature now.
There are two main ingredients of a \gls{mlp}: the regression model and the descriptors as inputs to the regression model.
For the construction of input descriptors, linearly complete basis functions for the atom-environments have been proposed \cite{Shapeev2016, drautz2019prb}.
For the regression model, linear regression \cite{Thompson2014jcp, Shapeev2016}, artificial \gls{nn} regression \cite{behler2007prl}, and kernel-based regression \cite{bartok2010prl} have all been proven to be feasible approaches.
The combination of equivariant (as opposed to invariant) constructions and message passing or graph \glspl{nn} \cite{batzner2022nc, Batatia2022} has also shown great potential in enhancing the regression accuracy of \glspl{mlp}, albeit at the cost of reduced computational efficiency and challenges in maintaining parallelism.

Despite the higher accuracy offered by \glspl{mlp}, there are still challenges for applying \glspl{mlp} in materials modeling, namely the relatively higher computational cost of many \glspl{mlp} compared to most conventional many-body potentials, and the absence of readily usable databases of \glspl{mlp} that cover a large number of elements and their compounds.
In some cases where an extensive database is available, one can use an available \gls{mlp} to study a specific problem, but in many cases, one has to train a new one or improve an existing one before being able to study the problem at hand.
In particular, there is no simple way to combine \glspl{mlp} for different elements to build \glspl{mlp} for their compounds or alloys.
This can lead to repeated efforts in the community and the case-by-case approach of developing \glspl{mlp} is neither optimal nor sustainable in the long run.
Regarding the computational cost of \glspl{mlp}, the \gls{nep} approach \cite{fan2021neuroevolution, fan2022jpcm, fan2022jcp} developed recently has been shown to yield excellent computational efficiency compared to other state-of-the-art methods, thanks to an optimization of the theoretical formalism and an efficient implementation in the \textsc{gpumd} package \cite{fan2017cpc}.
The \gls{nep} approach can reach computational speeds unprecedented for \glspl{mlp}, on par with empirical potentials, paving the way for the application of \glspl{mlp} to large-scale atomistic simulations. 

In this paper, we introduce a sustainable approach for the construction of \glspl{mlp}.
Although our approach can in principle be utilized to construct a comprehensive \gls{mlp} covering the entire periodic table, we have chosen a more focused task as a proof of concept.
Our objective is to develop a general-purpose \gls{nep} model encompassing 16 elemental metals and their alloys.
Previous attempts to create general-purpose \glspl{mlp} for numerous elements, or even the entire periodic table, have been initiated by researchers such as Takamoto \textit{et al.} \cite{takamoto2022cms, takamoto2022nc} and Chen and Ong \cite{chen2022ncs}.
These studies have introduced ``universal'' \glspl{mlp}, covering up to \num{45} elements \cite{takamoto2022nc} and \num{89} elements \cite{chen2022ncs}, respectively.
Despite being termed universal, these \glspl{mlp} have a rather limited application range and are orders of magnitude slower than \gls{eam} potentials.
General-purpose \glspl{mlp} have only been conclusively demonstrated for elemental matter such as Si \cite{bartok2018prx}, C \cite{rowe2020jcp}, Fe \cite{Jana2023prb}, and Pb \cite{kloppenburg2023jcp}.
For compounds comprising multiple chemical species, a special class of transition-metal oxides \cite{Artrith2017prb}, binary Sn alloys with a few metals \cite{Thorn2023pccp}, and Si--O \cite{erhard2024modelling} have been successfully modeled using \glspl{mlp}.
However, for metallic alloys, it remains a highly nontrivial task to construct a unified \gls{mlp} that can be reliably used for arbitrary chemical compositions.
Here, our goal is to construct a genuinely general-purpose \gls{mlp} for a diverse range of elements that matches the speed of \gls{eam} and surpasses it in the description of various physical properties. 

Apart from achieving high accuracy and efficiency for the unified \gls{nep} model, which we term \gls{unep1}, we also propose an efficient approach for constructing the training dataset.
Constructing a training dataset with all the possible chemical compositions is a formidable task.
Fortunately, the \gls{nep} descriptor parameters depend only on pairs of elements.
We will demonstrate that considering unaries and binaries alone for the training dataset is sufficient, yielding a \gls{nep} model that is transferable to systems with more components.
Using this route, we achieve a transferable \gls{unep1} model for 16 elemental metals (Ag, Al, Au, Cr, Cu, Mg, Mo, Ni, Pb, Pd, Pt, Ta, Ti, V, W, Zr) and their diverse alloys, with only about \num{100000} reference structures.
This accomplishment is evidenced by accurate predictions of formation energies across various test datasets comprising multi-component alloy systems, reproduction of experimentally observed chemical order and stable phases, and the generality and high efficiency of our \gls{unep1} model in large-scale \gls{md} simulations of mechanical deformation and primary radiation damage in MoTaVW refractory high-entropy alloys.

\section{Results}

\noindent{\textbf{A neural-network architecture for many-component systems.}}
Our starting point is the \gls{nep} approach as described in Ref.~\citenum{fan2022jcp}, called NEP3.
In this work, we introduce two crucial extensions to NEP3 designed specifically for many-component systems.
This extended approach will be called NEP4, which has been implemented in \textsc{gpumd}  during the course of this work and is available starting from version 3.8. 

We first briefly introduce NEP3 \cite{fan2022jcp}, which is a \gls{nn} potential that maps a descriptor vector $\mathbf{q}^i$ (with $N_\mathrm{des}$ components) of a central atom $i$ to its site energy $U_i$.
The total energy of a system of $N$ atoms is expressed as the sum of the site energies $U=\sum_{i=1}^N U^i$.
The \gls{ml} model is a fully connected feedforward \gls{nn} with a single hidden layer with $N_\mathrm{neu}$ neurons,
\begin{equation}
\label{equation:Ui_full}
U^i = \sum_{\mu=1}^{N_\mathrm{neu}}w^{(1)}_{\mu}\tanh\left(\sum_{\nu=1}^{N_\mathrm{des}} w^{(0)}_{\mu\nu} q^i_{\nu} - b^{(0)}_{\mu}\right) - b^{(1)},
\end{equation}
where $\tanh(x)$ is the activation function in the hidden layer, $\mathbf{w}^{(0)}$ is the connection weight matrix from the input layer (descriptor vector) to the hidden layer, $\mathbf{w}^{(1)}$ is the connection weight vector from the hidden layer to the output layer, $\mathbf{b}^{(0)}$ is the bias vector in the hidden layer, and $b^{(1)}$ is the bias in the output layer.
Denoting the weight and bias parameters in the \gls{nn} collectively as $\mathbf{w}$, we can formally express the site energy as 
\begin{equation}
\label{equation:Ui}
U^i = \mathcal{N}\left(\mathbf{w}; \mathbf{q}^i\right).
\end{equation}
The descriptor vector consists of a number of radial and angular components. In this work, we utilize up to five-body angular components.
For illustration purposes, we discuss the three-body angular components here.
Interested readers are referred to Ref.~\citenum{fan2022jcp} for the description of higher-order terms up to five-body angular components, which can help improve the model's completeness \cite{Pozdnyakov2021prl}.
A three-body angular descriptor component can be expressed as
\begin{equation}
\label{equation:qin}
q^i_{nl}
= \sum_{j\neq i} \sum_{k\neq i} g_{n}(r_{ij}) g_{n}(r_{ik}) P_l(\theta_{ijk}),
\end{equation}
where $n$ and $l$ represent the order of the radial and angular expansions, respectively.
Here, the summation runs over all neighbors of atom $i$ within a certain cutoff distance, $r_{ij}$ represents the distance between atoms $i$ and $j$, $\theta_{ijk}$ is the angle for the triplet $(ijk)$ with $i$ being the central atom, and $P_l(x)$ is the Legendre polynomial of order $l$.
The functions $g_n(r_{ij})$ depend solely on the distance $r_{ij}$ and are therefore referred to as radial functions.
These radial functions are defined as linear combinations of a number of  basis functions:
\begin{align}
g_n(r_{ij}) &= \sum_{k}c^{IJ}_{nk} f_k(r_{ij}).
\label{equation:g_n}
\end{align}
The basis functions $f_k$ are constructed based on Chebyshev polynomials and a cutoff function, ensuring both formal completeness and smoothness.
Explicit expressions for these functions can be found in Ref.~\citenum{fan2022jcp}.
The expansion coefficients $c_{nk}^{IJ}$ depend on $n$ and $k$ and also on the types (denoted as capitals $I$ and $J$) of atoms $i$ and $j$.
Due to the summation over neighbors, the descriptor components defined above are invariant with respect to permutation of atoms of the same type.
More importantly, these coefficients are treated as trainable parameters \cite{fan2022jpcm}, which is crucial for efficiently differentiating different atom pairs contributing to the descriptor. 

\begin{figure}
\includegraphics[width=1\columnwidth]{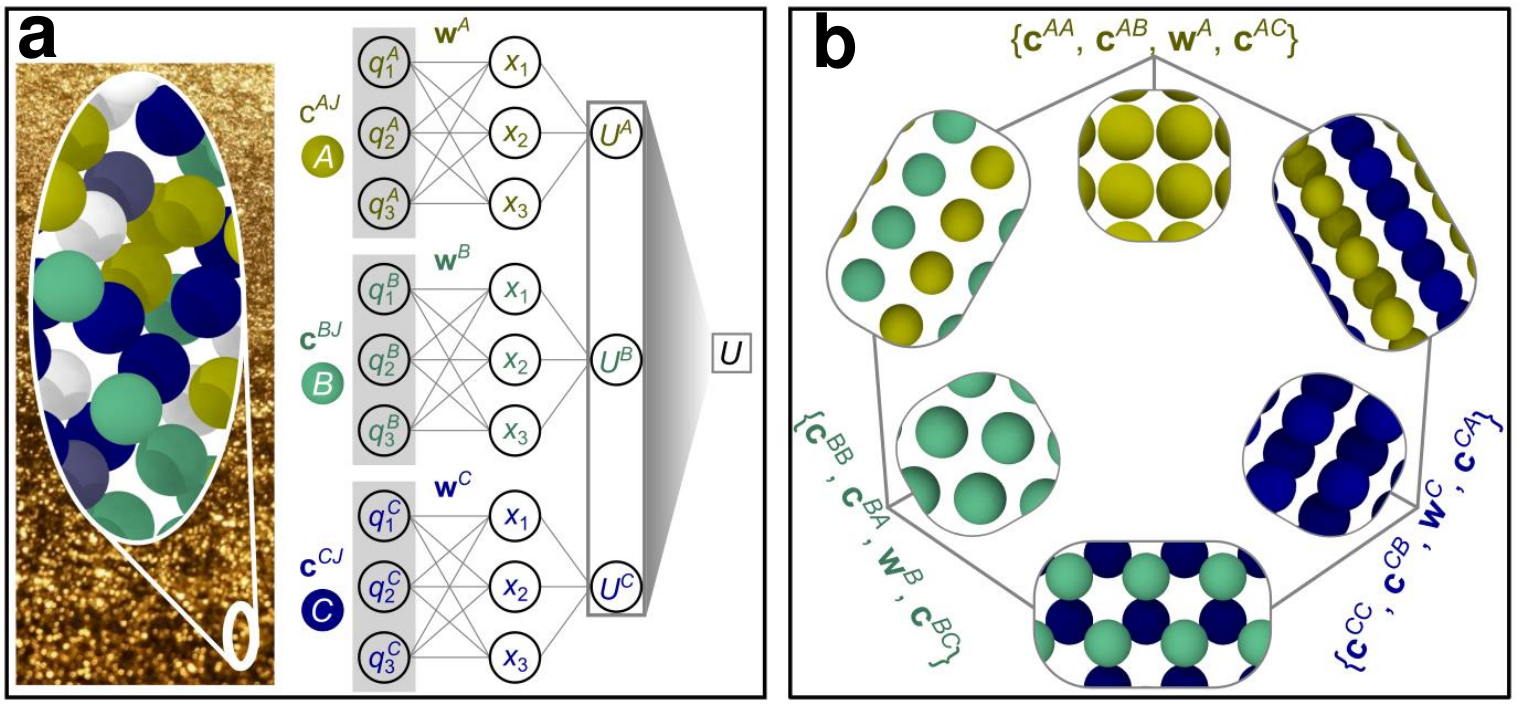}
\caption{
    \textbf{a} Schematic illustration of the architecture of the NEP4 model with distinct sets of \gls{nn} parameters for different atom types.
    For a central atom of type $A$, the descriptor involves the $\mathbf{c}^{AJ}$ parameters ($J$ can be of any type), while the weight and bias parameters $\mathbf{w}^A$ are specific for type $A$.
    Similar rules apply to the central atoms of other types.
    The total energy $U$ is the sum of the site energies for all the atoms in a given structure.
    By contrast, in NEP3 all atom types share a common set of \gls{nn} parameters $\mathbf{w}$, which restricts the regression capacity.
    \textbf{b} Schematic illustration of the multi-loss evolutionary training algorithm.
    For a 3-component system, the optimization of the parameters related to atom type $A$ (including $\mathbf{w}^A$, $\mathbf{c}^{AA}$, $\mathbf{c}^{AB}$, and $\mathbf{c}^{AC}$) is only driven by a loss function defined using the structures with the chemical compositions of $A$, $AB$, and $AC$.
    In the conventional evolutionary algorithm, which is used in NEP3, a single loss function is used to optimize all parameters, which is less effective for training general-purpose models for many-component systems. 
}
\label{fig:nep4}
\end{figure}

While the descriptor parameters $\{\mathbf{c}^{IJ}\}$ depend on the atom types (species), the \gls{nn} parameters $\mathbf{w}$ in NEP3 are the same for all the atom types.
Therefore, as the number of atom types increases, the regression capacity of the \gls{nn} model for each atom type decreases.
To keep a constant regression capacity per atom type, in the present work, we employ different sets of \gls{nn} parameters $\mathbf{w}^I$ for each atom type $I$.
While this increases the total number of trainable parameters, it \emph{does not} significantly increase the computational cost during \gls{md} simulations with the trained model, because it only involves a selection of the correct set of \gls{nn} parameters for a given atom.
With the extension, the site energy can be expressed as
\begin{equation}
\label{equation:Ui_nep4}
U^i = \mathcal{N}\left(\mathbf{w}^I; \mathbf{q}^i(\{\mathbf{c}^{IJ}\})\right),
\end{equation}
which constitutes the NEP4 model introduced in this work (\autoref{fig:nep4}a).

\vspace{0.5cm}

\noindent{\textbf{A multiple-loss evolutionary training algorithm for many-component systems.}}
While the increase in the number of trainable parameters does not significantly affect the inference speed, it considerably increases the number of iterations required for training, particularly with the approach used for NEP3.
It turns out that the training algorithm must be modified to achieve better performance for many-element systems.
For training \gls{nep} models we use the \gls{snes} approach \cite{Schaul2011}, which is a powerful black-box optimization algorithm that is particularly suitable for problems with many possible solutions \cite{wierstra2014jmlr}.
It maintains a mean value and a variance for each trainable parameter that are updated according to the rank of a population of solutions.
The rank is determined according to the loss function to be minimized.
The loss function $L$ is constructed using predicted and reference data for energies, forces, and virials, and is a function of the trainable parameters, i.e.,
\begin{equation}
L = L\left(\{\mathbf{w}^I\}; \{\mathbf{c}^{IJ}\}\right).
\label{equation:L}
\end{equation}

The rank (or ``fitness'')  is of crucial importance in evolutionary algorithms, as it determines the relative weight of a solution in the population.
However, using a single loss function can lead to ambiguity in rank assignment:
Even if the total loss of solution X is smaller than that of solution Y, it does not guarantee that solution X is more accurate for all the subsystems in a many-element system.
For example, solution X might offer higher accuracy for Au systems but lower accuracy for Ag systems.
To account for this observation we define multiple loss functions for many-element systems.
Since we are concerned with alloys, we cannot define a set of loss functions that have no common terms at all, but we can make a definition that minimizes the common parts.
Naturally, we define the loss function for element $I$ as the parts in Eq.~\eqref{equation:L} that are contributed by structures containing element $I$.
For illustration, consider an explicit example with three elements, denoted $A$, $B$, and $C$, respectively.
The loss function for element $A$ can be calculated by considering the chemical compositions $A$, $AB$, and $AC$ only, excluding $B$, $C$, and $BC$.
This loss function is used when training the parameters related to element $A$, which are $\mathbf{w}^A$, $\mathbf{c}^{AA}$, $\mathbf{c}^{AB}$, and $\mathbf{c}^{AC}$ (\autoref{fig:nep4}b).
Using this multi-loss evolutionary algorithm, the training converges much faster than using a single-loss function.
The efficiency improvement in training becomes more significant with an increasing number of elements, and is crucial for being able to develop models such as \gls{unep1}.

\vspace{0.5cm}

\noindent{\textbf{Construction of training data for many-component systems based on chemical generalizability.}}
The chemical space for 16 elements consists of $2^{16}-1=\num{65535}$ chemical combinations, including 16 unaries, 120 binaries, 560 ternaries, etc. 
It is formidable to construct a training dataset by enumerating all the possible chemical combinations. 
Fortunately, leveraging the construction of the radial functions in terms of linear combinations of basis functions provides a solution. 
The descriptor values for a given configuration of $n$-component ($n>2$) systems fall within the range spanned by those of the 1-component and 2-component systems derived from the same configuration by element substitution.
Given the interpolation capabilities of \glspl{nn}, a \gls{nep} model trained using 1-component and 2-component structures is expected to predict the behavior of $n$-component ($n>2$) systems reasonably well. 
Therefore, our training dataset focused only on unary and binary systems.

For each unary or binary system, we constructed an initial training dataset with a few hundred structures. 
These structures included small cells with position and/or cell perturbations, cells with one to a few vacancies, cells with surfaces and various defects (such as grain boundaries) taken from the Materials Project \cite{Jain2013aplm} and the Open Quantum Materials Database \cite{kirklin2015npjcm}, cells sampled from \gls{md} simulations based on an \gls{eam} potential \cite{zhou2004prb} at various temperature (up to \qty{5000}{\kelvin}) and pressure conditions including highly deformed structures (see Methods for details).
There are initially about \num{60000} structures in total for the 16 metals and their binary alloys. 
In spite of its seemingly modest size, this training dataset is remarkably diverse in configuration space.
Reference data (energy, force, and virial) for the structures were generated via \gls{dft} calculations using the VASP package (see Methods for details). 

The diversity of the initial training dataset ensured a robust initial \gls{nep} model that could be used to run \gls{md} and \gls{mcmd} simulations at various thermodynamic conditions. 
From diverse \gls{md} and \gls{mcmd} trajectories generated by the initial \gls{nep} model, structures (still unary and binary only) were sampled and labeled using \gls{dft} calculations. 
Those with relatively large errors (\gls{nep} versus \gls{dft}) were identified and incorporated into the training set. 
This iterative process was repeated a few times until no large errors could be detected. 
This active-learning scheme, while simple, proved to be highly effective.
The final training dataset contains \num{105464} structures and \num{6886241} atoms in total. 
The \gls{dft} calculations for these structures required about six million CPU hours.

\begin{figure*}
\includegraphics[width=2\columnwidth]{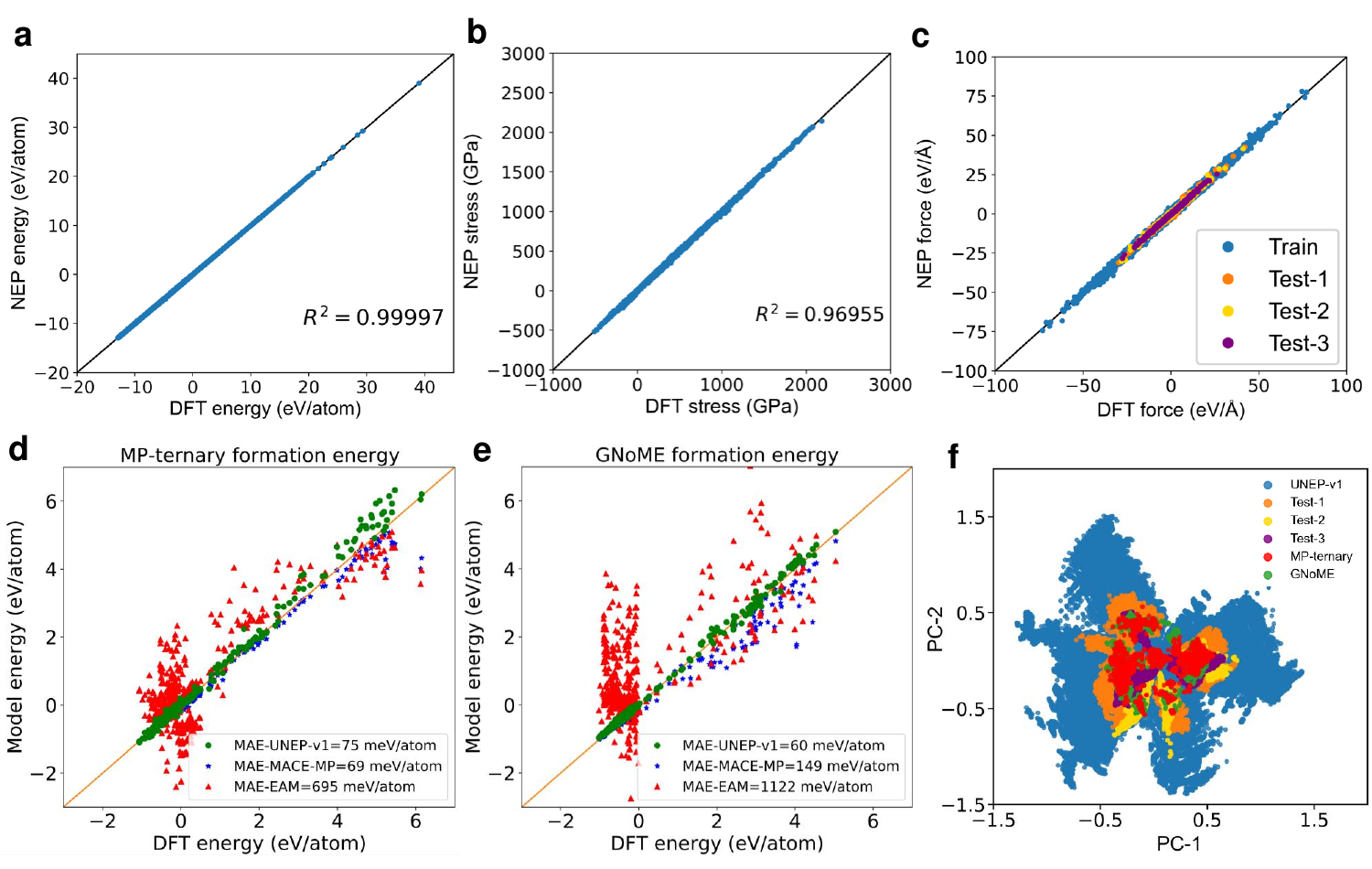}
\caption{
    \textbf{a}--\textbf{c} Parity plots for energy, stress, and force comparing \gls{dft} reference data and \gls{unep1} predictions for the whole training dataset.
    In \textbf{c}, there are three test datasets containing $n$-component ($n\geq 3$) structures, including one with up to 13 components (Ag, Au, Cr, Cu, Mo, Ni, Pd, Pt, Ta, Ti, V, W, Zr) taken from Lopanitsyna \textit{et al.} \cite{Lopanitsyna2023PRM} (labeled Test-1), one with up to four components (Mo, Ta, V, W) from Byggm\"astar \textit{et al.} \cite{Byggmastar2022prm} (labeled Test-2), and one with up to three components (Pd, Cu, Ni) from Zhao \textit{et al.} \cite{zhao2023md} (labeled Test-3).
     \textbf{d}--\textbf{e} Parity plots for formation energies comparing \gls{dft} reference data and predictions from \gls{unep1}, MACE-MP-0 (medium model) \cite{batatia2024foundation}, and \gls{eam} \cite{zhou2004prb}, for structures from the Materials Project (MP-ternary)  \cite{Jain2013aplm} and the GNoME paper \cite{merchant2023nature}.
    \textbf{f} Distribution of the training dataset (this work, \gls{unep1}, comprising 1-component to 2-component systems, blue) and various test datasets, including Test-1 (up to 13-component systems, orange) \cite{Lopanitsyna2023PRM}, Test-2 (up to 4-component systems, yellow) \cite{Byggmastar2022prm}, Test-3 (up to 3-component systems, purple) \cite{zhao2023md}, MP-ternary alloys (3-component systems, red) \cite{Jain2013aplm}, and GNoME dataset (2-component to 5-component systems, green) \cite{merchant2023nature}, in the 2D principal component (PC) space of the descriptor.
}
\label{fig:parity-train}
\end{figure*}

\vspace{0.5cm}

\noindent{\textbf{Training and testing results.}}
Using the refined training dataset, we trained a \gls{nep} model (see Method and Supplementary Note for details on the hyperparameters) using the NEP4 approach as described above.
We refer to this \gls{nep} model as \gls{unep1}, which represents the first attempt at constructing a unified \gls{nep} model for many elements.

The parity plots for energy, force, and stress affirm the high accuracy of this \gls{unep1} model (\autoref{fig:parity-train}a-c). 
Despite the large ranges of the three quantities, their \glspl{rmse} are relatively small, at \qty{17.1}{\milli\electronvolt\per\atom}, \qty{172}{\milli\electronvolt\per\angstrom}, and \qty{1.16}{\GPa}, respectively.

To validate the force accuracy of our \gls{unep1} model we consider here three public datasets.
Although the public datasets were not computed using exactly the same \gls{dft} settings as used for generating the \gls{unep1} training data, the resulting differences in force values are marginal (of the order of a few \unit{\milli\electronvolt\per\angstrom}) and are much smaller than the force \gls{rmse} achieved by \gls{unep1} (\autoref{fig:parity-train}c).
The comparison moreover shows that the \gls{unep1} model trained against 1-component and 2-component structures also performs very well for 3-component \cite{zhao2023md}, 4-component \cite{Byggmastar2022prm}, and 13-component \cite{Lopanitsyna2023PRM} structures extracted from the datasets in the previous studies \cite{zhao2023md, Byggmastar2022prm, Lopanitsyna2023PRM}.
The testing \glspl{rmse} of the \gls{unep1} model for these three datasets are respectively \qty{76}{\milli\electronvolt\per\angstrom}, \qty{196}{\milli\electronvolt\per\angstrom}, and \qty{269}{\milli\electronvolt\per\angstrom}, which are comparable to those reported as training \glspl{rmse} in the original publications \cite{zhao2023md, Byggmastar2022prm, Lopanitsyna2023PRM}.

To validate the energy accuracy of our \gls{unep1} model we utilize two public datasets, including all the relevant 3-component structures in the Materials Project database \cite{Jain2013aplm} and the structures predicted using the GNoME approach \cite{merchant2023nature} ranging from 2-component to 5-component systems with force components less than \qty{80}{\electronvolt\per\angstrom}.
We calculate the formation energies using \gls{dft}, an \gls{eam} potential \cite{zhou2004prb}, a foundation model named MACE-MP-0 (medium version) \cite{batatia2024foundation}, and our \gls{unep1} model, where the reference energy for each species is based on the most stable allotrope.
For the two datasets, the \glspl{mae} of our \gls{unep1} model compared to \gls{dft} calculations are \qty{75}{\milli\electronvolt\per\atom} and \qty{60}{\milli\electronvolt\per\atom}, respectively (\autoref{fig:parity-train}d and \autoref{fig:parity-train}e).
In contrast, the corresponding values from the \gls{eam} potential are \qty{695}{\milli\electronvolt\per\atom} and \qty{1122}{\milli\electronvolt\per\atom}, respectively, and thus about one order of magnitude larger.
For the Materials Project dataset, which MACE-MP-0 has been trained on while \gls{unep1} has not, MACE-MP-0 is slightly more accurate.
However, for the GNoME dataset, on which neither model has been trained, \gls{unep1} demonstrates notably better accuracy.

Besides the Materials Project and GNoME datasets, Figures S1--S3 present parity plots for formation energies and forces predicted by \gls{unep1}, \gls{eam}, and MACE-MP-0 compared to DFT for the three test datasets \cite{zhao2023md, Byggmastar2022prm, Lopanitsyna2023PRM}.
Figures S4--S17 show the formation energies, comparing \gls{unep1}, \gls{eam}, and DFT for the equation of state curves (for alloys), heating, compressing, and stretching processes with 1 to 5-component materials in various crystalline structures, including \gls{fcc}, \gls{bcc}, \gls{hcp}, and metallic glasses.
The results altogether clearly demonstrate the superior accuracy of \gls{unep1} over \gls{eam} and confirm the excellent generalizability of our \gls{unep1} model from the 1- and 2-component structures included in the training dataset to unseen multi-component structures.

As a further test, we trained a \gls{nep} model by including relevant $n$-component ($n \geq 3$) structures from the Open Quantum Materials Database database \cite{kirklin2015npjcm}.
The \glspl{rmse} for the three public datasets \cite{zhao2023md, Byggmastar2022prm, Lopanitsyna2023PRM} obtained using this \gls{nep} model are only marginally improved compared to \gls{unep1}, which demonstrates that our training dataset with $n$-component ($n \leq 2$) structures is already sufficient for training a general-purpose \gls{nep} model for all the considered elements and their alloys.

As mentioned earlier, our approach to training data generation relies on the chemical generalizability embedded in the radial functions Eq.~\eqref{equation:g_n}.
This feature is illustrated by a principal component analysis of the descriptor space (\autoref{fig:parity-train}f), which shows that the various $n$-component ($n\geq 3$) structures fall comfortably within the space spanned by the 1-component and 2-component training structures. 

\vspace{0.5cm}

\begin{figure}
\centering
\includegraphics[width=\columnwidth]{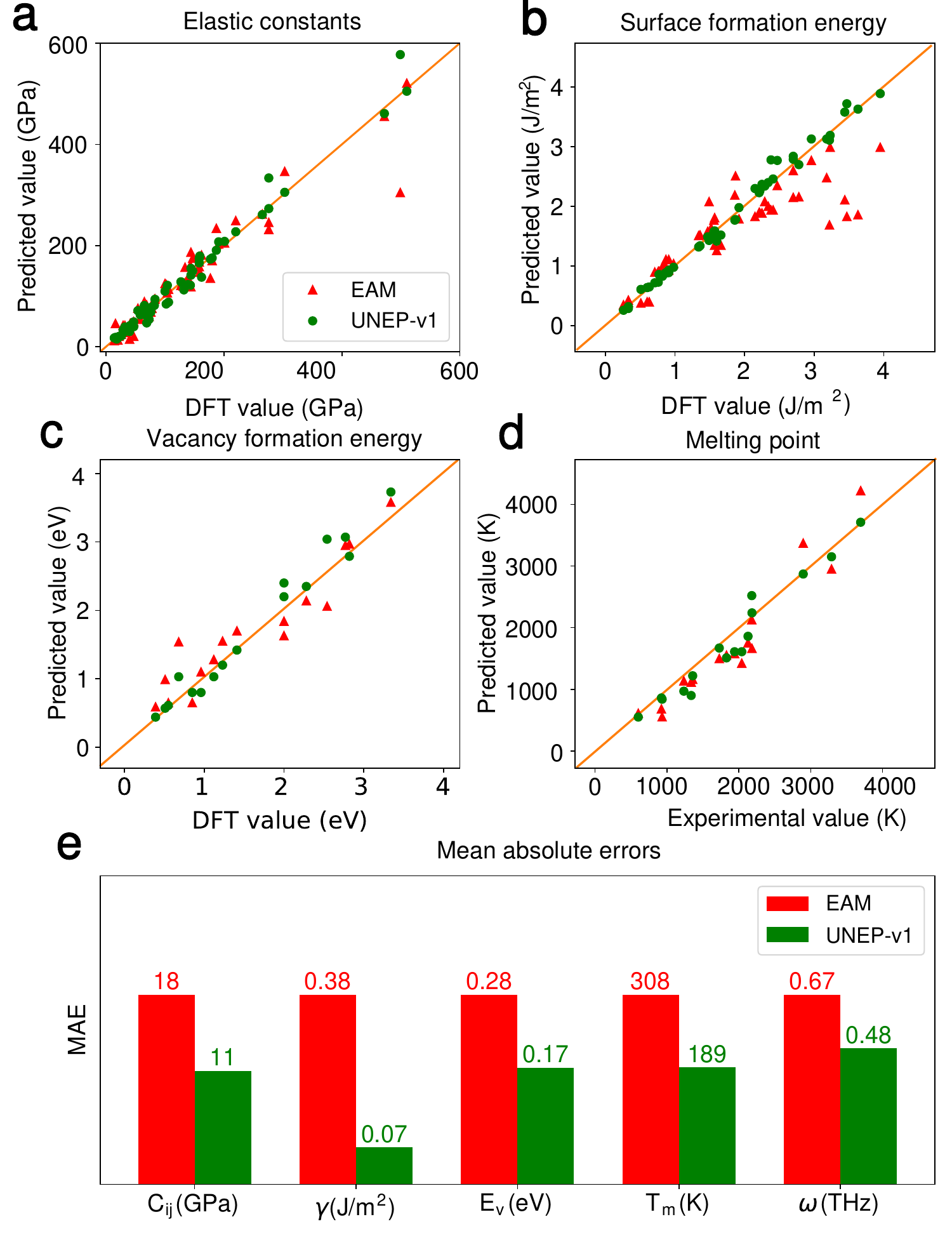}
\caption{
    \textbf{a}--\textbf{d} Elastic constants $C_{ij}$, formation energies $\gamma$ for $\left\{111\right\}$, $\left\{110\right\}$, and $\left\{100\right\}$ surfaces, mono-vacancy formation energies $E_{\rm v}$, and melting points $T_{\rm m}$ as predicted by the \gls{eam} potential \cite{zhou2004prb} and the \gls{unep1} model compared to \gls{dft} or experimental values for the 16 elements.
    \textbf{e} \Acrfullpl{mae} for the above four quantities as well as the phonon frequency $\omega$ for \gls{eam} and \gls{unep1} models with respect to reference data from \gls{dft} calculations and experiment.
}
\label{fig:mae}
\end{figure}

\noindent\textbf{Evaluation of basic physical properties for the 16 metal elements.}
After having confirmed the high training accuracy of the \gls{unep1} model for 1-component and 2-component systems, and its high testing accuracy for systems with multiple components, we conducted an extensive evaluation of the \gls{unep1} model beyond \glspl{rmse}, focusing on various physical properties (see Methods for details on the calculations).
Elastic constants $C_{ij}$, surface formation energies $\gamma$, mono-vacancy formation energies $E_{\rm v}$, melting points $T_{\rm m}$, and phonon dispersion relations were calculated for all 16 elements, using both the \gls{unep1} model and an \gls{eam} potential \cite{zhou2004prb}. 
While there are recent and possibly more accurate \gls{eam} potentials \cite{Sheng2011prb} for a limited subset of species considered here, we have consistently opted for the widely used \gls{eam} potential developed by Zhou \textit{et al.} \cite{zhou2004prb} because it supports all the 16 species and their alloys. 
Detailed results for phonon dispersion relations are presented in Figs. S18--S20, while other physical properties are listed in Tables S1--S4.
Figures \ref{fig:mae}a--d show the parity plots comparing predictions of various basic properties from \gls{eam} and \gls{unep1} against \gls{dft} calculations or experimental values.
The \gls{eam} predictions have some outliers, especially in the case of the surface formation energies, while the \gls{unep1} predictions do not show any notable discrepancies. 

The \glspl{mae} for all the evaluated quantities calculated by averaging the absolute error between predicted (\gls{eam} or \gls{unep1}) and reference values (\gls{dft} or experimental) over all 16 elements are presented in \autoref{fig:mae}e.
\gls{unep1} consistently outperforms the \gls{eam} potential for all physical properties, and demonstrates a significant advantage in predicting surface formation energies, elastic constants, and vacancy formation energies.

\vspace{0.5cm}

\begin{figure}
    \centering    
    \includegraphics[width=1.0\columnwidth]{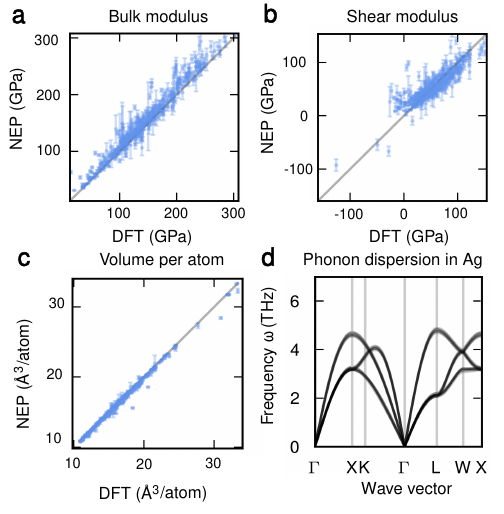}
    \caption{
        \textbf{a}--\textbf{c} Parity plots for \gls{nep} model versus \gls{dft} data for bulk modulus, shear modulus, and equilibrium volume for roughly 400 different alloys of the 16 elements, with error bars computed as the standard deviation in the predicted properties over an ensemble of eight \gls{nep} models.
        The structures and reference \gls{dft} data were taken from the Materials Project \cite{Jain2013aplm}.
        \textbf{d} Phonon dispersion relations for \gls{fcc} Ag calculated by averaging of all models in the ensemble.
    }
    \label{fig:ensemble-properties}
\end{figure}

We have additionally trained an ensemble of eight \gls{nep} models using different sets of training hyperparameters, and compared the predictions for bulk and shear moduli as well as the equilibrium volume for the ensemble to \gls{dft} reference data to estimate the uncertainty in the model predictions (\autoref{fig:ensemble-properties}a--c).
Generally, the deviations in the predictions across the ensemble are small, and mostly agree well with the reference data.
As a further illustration, we estimated the uncertainty in the phonon dispersion for Ag (\autoref{fig:ensemble-properties}d), illustrating the very small uncertainty throughout the entire Brillouin zone.

\vspace{0.5cm}

\noindent\textbf{Computational performance.}
The computational efficiency of a \gls{mlp} is crucial for its effective applications in large-scale \gls{md} simulations. 
Here, the \gls{unep1} model as implemented in \textsc{gpumd} exhibits excellent computational performance (\autoref{table:speed}).
Using a single Nvidia A100 GPU, \gls{unep1} can reach a simulation size of about 14 million atoms and a computational speed of \qty{2.4e7}{\atomstepsecond}, which is only a few times lower than that for the \gls{eam} potential (\qty{11e7}{\atomstepsecond}) as implemented in \textsc{lammps} \cite{thompson2022cpc} using the same hardware.
To reach even larger simulation sizes, we implemented a multi-GPU version of \gls{nep} that can effectively use the computational power of all the GPUs available on a computational node.
With only 8 A100 GPUs, we can reach a simulation size of 100 million atoms, achieving much higher computational efficiency than either the \gls{dp} (thousands of Nvidia V100 GPUs) \cite{jia2020GB, Guo2022DP} or Allegro (128 A100 GPUs) approaches \cite{musaelian2023nc}. 

\begin{table}[thb]
\centering
\setlength{\tabcolsep}{2Mm}
\caption{
    Computational performance of \gls{unep1} in comparison with \gls{dp} \cite{jia2020GB, Guo2022DP}, Allegro \cite{musaelian2023nc}, and \gls{eam} models (using the GPU package of \textsc{lammps} \cite{thompson2022cpc}) for large-scale \gls{md} simulations of typical metals.
    The speed is given in units of \qty{e3}{\atomstepsecond\per\gpu}.
    The \gls{dp} results were obtained on V100 GPUs.
    All other data were generated using A100 GPUs, which offer approximately twice the computational performance of a V100 GPU for this kind of computation.
}
\label{table:speed}
\begin{tabular}{lrrr}
\hline
\hline
Model-Element & \# atoms & \# GPUs  & Speed \\
\hline
\gls{dp}-Cu \cite{jia2020GB}      & \num{127e6}  & \num{27300} & \num{45} \\ 
\gls{dp}-Cu \cite{Guo2022DP}      & \num{3400e6} & \num{27300} & \num{330} \\[3pt]
Allegro-Ag \cite{musaelian2023nc} & \num{100e6}  & 128         & \num{2600} \\[3pt]
\gls{eam}-Cu                      & \num{23e6}   & 1           & \num{110000} \\
\gls{eam}-Cu                      & \num{100e6}  & 4           & \num{49300} \\[3pt]
\gls{unep1}-Cu                    & \num{14e6}   & 1           & \num{23500} \\
\gls{unep1}-Cu                    & \num{100e6}  & 8           & \num{18800} \\ 
\gls{unep1}-Ag                    & \num{100e6}  & 8           & \num{17200} \\
\hline
\hline
\end{tabular}
\end{table}

With 8 A100 GPUs, the overall computational speed of \gls{unep1} is about \qty{1.5e8}{\atomstepsecond}.
The parallel efficiency relative to ideal scaling for \gls{unep1} with 8 A100 GPUs is 80\%, while it is only about 50\% for \gls{eam} with 4 A100 GPUs. 
The speed per GPU achieved by \gls{unep1} is significantly higher than those for the \gls{dp} \cite{jia2020GB, Guo2022DP} and Allegro approaches \cite{musaelian2023nc}.
The excellent computational speed of \gls{unep1} allows us to tackle challenging problems in \glspl{mpea} as discussed below.

\vspace{0.5cm}

\begin{figure*}
\centering
\includegraphics[width=2\columnwidth]{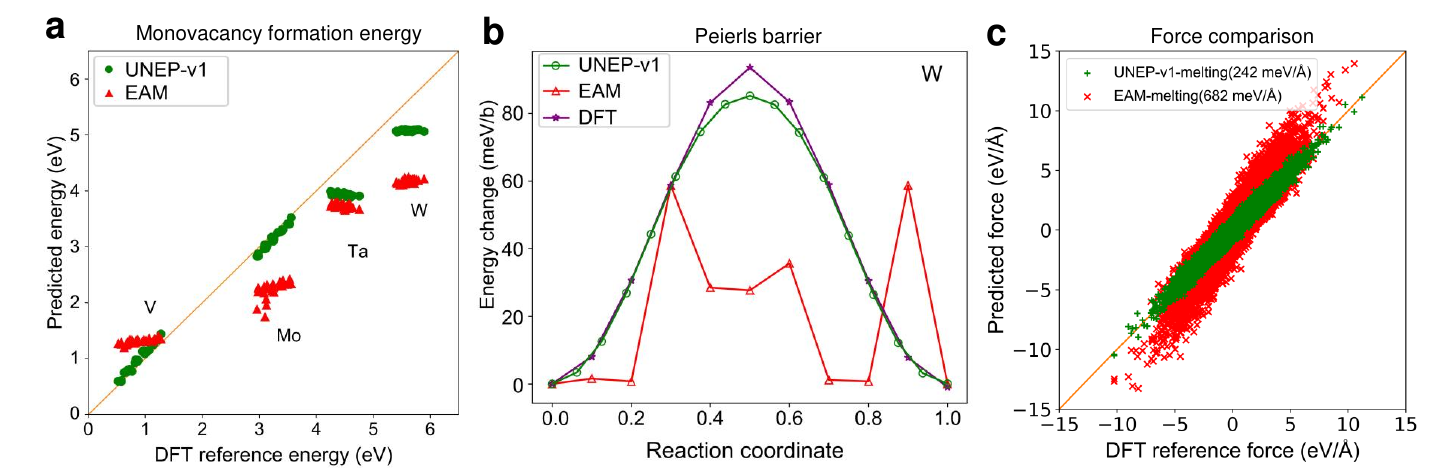}
\caption{
    \textbf{a} Mono-vacancy formation energies from \gls{unep1} and \gls{eam} \cite{zhou2004prb} compared to \gls{dft} data for an equimolar MoTaVW alloy with 128 atoms sampled from \gls{mcmd} simulations.
    \textbf{b} Peierls barrier for $1/2\langle 111\rangle$ screw dislocation migration in elemental W (see Fig.~S21 for the other three species).
    \textbf{c} Comparisons of \gls{unep1}, \gls{eam} \cite{zhou2004prb}, and \gls{dft} results for equimolar MoTaVW alloys sampled from \gls{md} simulations using 256-atom supercells for a melting process from 10 to \qty{5000}{\kelvin} during \qty{10}{\nano\second}. \gls{unep1} shows much better predictions than \gls{eam}, with a much smaller force \gls{rmse} as indicated in the legend (see Fig. S22 for similar comparisons for deformation processes).
}
\label{fig:MoTaVW-test}
\end{figure*}

\noindent\textbf{Application to plasticity of multi-principal element alloys.}
Refractory \glspl{mpea} have emerged as materials for high-temperature applications, crystallizing typically in the \gls{bcc} solid solution phase.
These alloys exhibit exceptional properties  such as high ductility and mechanical strength at ultra-high temperature \cite{Senkov2019acta, George2019nrm, Coury2019acta, Shi2019nc} as well as impressive irradiation resistance \cite{Atwani2019sciadv, Atwani2023nc}.
However, their ductility at room temperature is limited \cite{Senkov2018jmr, Couzinie2019mc}.
Recent experimental observations in alloys such as HfNbTaTiZr have revealed the presence of numerous straight screw dislocations and a substantial amount of dislocation debris \cite{Couzinie2019mc, Lilensten2018acta}, consistent with known behavior in \gls{bcc} metals \cite{pure-metal}.
Recent \gls{md} simulations have also indicated the possible crucial role of dislocation in the plastic flow of \glspl{mpea} \cite{Li2020npj, Yin2021nc, Zheng2023npj}.
Despite these insights, the complex structural and mechanical properties of \glspl{mpea} remain incompletely understood.
Here, atomistic simulations employing accurate and efficient \glspl{mlp} can provide further insights into the intricate behavior of these materials.
Although there are a few available \glspl{mlp} limited to specific alloys \cite{Li2020npj, Yin2021nc, Zheng2023npj}, a comprehensive general-purpose potential model capable of encompassing a wide range of elements and their alloys, providing both high efficiency and accuracy and enabling large-scale (up to millions of atoms) \gls{md} simulations of \gls{bcc} \glspl{mpea}, is still lacking.

The \gls{unep1} model developed in this work emerges as a promising solution, enabling large-scale \gls{md} simulations of \glspl{mpea} with an accuracy superior to existing models while still achieving very high computational efficiency.
To demonstrate its effectiveness in this context, we investigated the mechanism of plastic deformation of a MoTaVW alloy under compression.
Our evaluation of the \gls{unep1} model involved comprehensive tests, including checking the vacancy formation energies (\autoref{fig:MoTaVW-test}a) in equimolar MoTaVW alloys, Peierls barriers for the $1/2\langle 111\rangle$ screw dislocation (\autoref{fig:MoTaVW-test}b) in elemental systems as well as atomic forces in melting (\autoref{fig:MoTaVW-test}c), compression and tensile stretching processes (Fig. S22) of equimolar MoTaVW alloys.
The results illustrate the superior performance of \gls{unep1} compared to \gls{eam} potentials and its suitability for studying structural and mechanical properties in large-scale \gls{md} simulations. 

\begin{figure}
    \includegraphics[width=1\columnwidth]{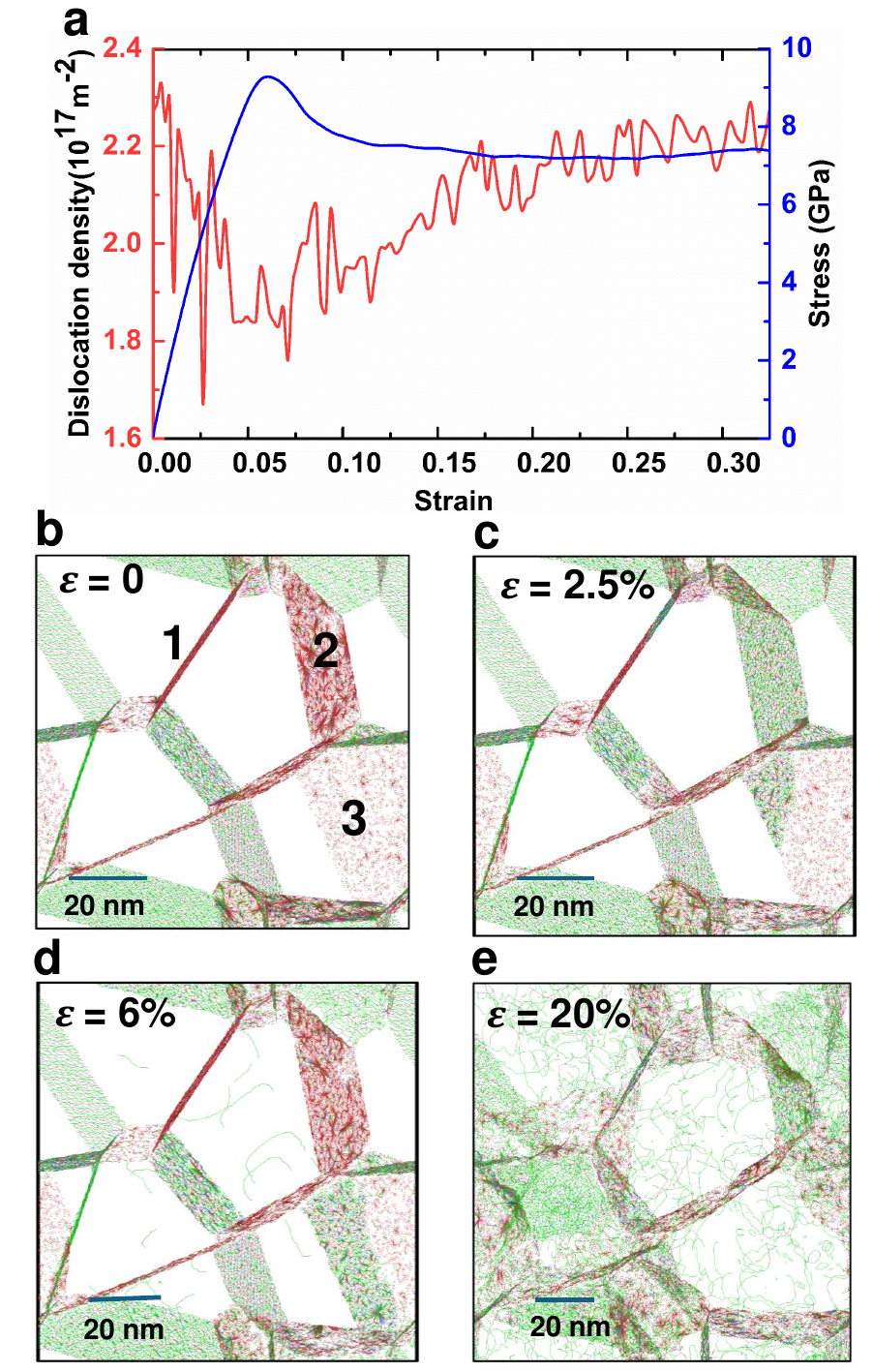}
    \caption{
        Dislocation density as a function of compressive strain for equimolar polycrystalline MoTaVW alloys containing 12 grains with 100 million atoms at \qty{300}{\kelvin}.
        \textbf{a} Strain-induced dislocation density and stress.
        \textbf{b}--\textbf{e} Distributions of dislocation in \qty{20}{\nano\meter} thick slices at strains of (\textbf{b}) $\epsilon=0\%$, (\textbf{c}) $ \epsilon=2.5\%$, (\textbf{d}) $ \epsilon=6\%$, and (\textbf{e}) $ \epsilon=20\%$, respectively.
        Grain boundaries are labeled by numbers for reference.
        The compressed direction is perpendicular to the plane of view.
        The $1/2\langle 111 \rangle$ dislocations are depicted in green, while other dislocations are shown in red.
    }
    \label{fig:strainedDislocation}
\end{figure}

After having confirmed the accuracy, efficiency, and reliability of our \gls{nep} model, we modeled an equimolar \gls{bcc} polycrystalline MoTaVW system containing \num{100205176} atoms and conducted \gls{md} simulations to investigate changes in dislocation density under compression.
These simulations involved compressive deformation at a strain rate of \qty{4.2e8}{\per\second} (see Methods for simulation details).
The dislocation density decreases during the elastic stage, reaches a minimum at the yield strain $\epsilon=6\%$, and gradually returns to the original level due to enhanced densification (\autoref{fig:strainedDislocation}a).
The dislocation density plateaus for large strains ($\epsilon\geq16\%$), consistent with the behavior observed in \gls{bcc} Ta \cite{Zepeda2017nature}.
It is noteworthy that stress-strain response and dislocation density exhibit contrasting trends under compression.  

To gain deeper insight into the plastic deformation mechanisms, we extracted the distribution of the dislocation density in snapshots of the polycrystalline MoTaVW system at selected strains (\autoref{fig:strainedDislocation}b--e).
Notably, all dislocations are confined to grain boundaries of the polycrystalline system under compression, and this pattern remains unchanged throughout the linear response (``elastic'') region of the stress-strain curve  (\autoref{fig:strainedDislocation}b--c).
It is worth noting that dislocations transform from other types (labeled 1 and 2 in \autoref{fig:strainedDislocation}b) to $1/2\langle 111 \rangle$ ones (\autoref{fig:strainedDislocation}c) in the elastic region ($0-2.5$\%), and recover back at the yield strain of 6\% (\autoref{fig:strainedDislocation}d).
Subsequently, during the plastic stage (\autoref{fig:strainedDislocation}d--e), some of the grain boundaries begin to emit, slip, and pin dislocations into the grains along with boundary movement.
This finding demonstrates the significant impacts of boundary stability on the hardness of \glspl{mpea}, as previously observed in the study of a NiMo alloy \cite{Hu2017science}.

Through 100-million-atom large-scale \gls{md} simulations, we have thus illuminated the intricate details of plastic deformation, shedding light on dislocation behavior in grain boundaries.
This application of our \gls{unep1} model to the plasticity of \glspl{mpea}, exemplified by the MoTaVW alloy, is an important demonstration for the generality and high computational efficiency of our approach.

\vspace{0.5cm}

\noindent{\textbf{Application to primary radiation damage in \glspl{mpea}.}}
Next, we demonstrate the versatility of the \gls{unep1} model through large-scale \gls{md} simulations of primary radiation damage in \glspl{mpea}, using again the MoTaVW alloy system for illustration (see Methods for details).
Notably, these simulations set a new benchmark with a record-breaking size of 16 million atoms for \glspl{mlp} in this specific type of simulations.
Here, in order to accurately describe interactions at extremely short distances where large forces are at play, we incorporated a two-body \gls{zbl} potential \cite{Ziegler1985} to train a combined \gls{nep}-\gls{zbl} model \cite{Liu2023prb}.
The \gls{unep1} part and the \gls{zbl} part are smoothly connected in the range between \num{1} and \qty{2}{\angstrom}.
Above \qty{2}{\angstrom}, only the \gls{unep1} part is active, while below \qty{1}{\angstrom}, the \gls{zbl} part dominates.
Illustrative examples demonstrating the seamless connection between \gls{unep1} and \gls{zbl} for Al and W dimers are presented in Fig.~S23.

\begin{figure}
\includegraphics[scale=0.4]{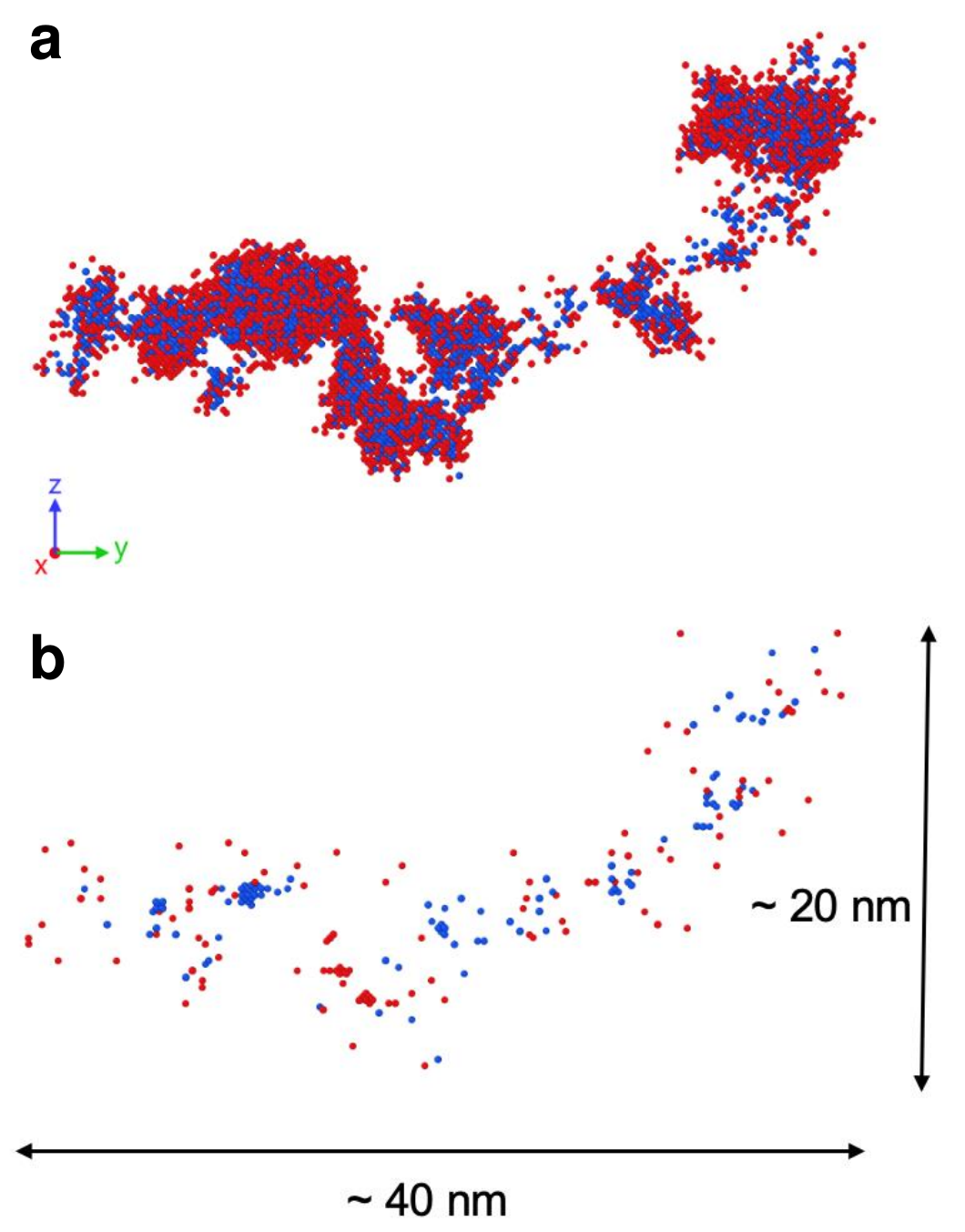}
\caption{
    Defect snapshots of a cascade in a MoTaVW alloy at (\textbf{a}) the peak damage state (at about \qty{0.6}{\pico\second}) and (\textbf{b}) the final damage state (at \qty{140}{\pico\second}).
    The red and blue dots represent interstitial atoms and vacancies, respectively.
}
\label{fig:cascade}
\end{figure}

Figure \ref{fig:cascade}a shows the defect snapshot of the peak-damage state formed at about \qty{0.6}{\pico\second} with a primary knock-on atom energy of \qty{100}{\kilo\electronvolt}.
The defect distribution stabilizes after a few tens of ps.
Figure \ref{fig:cascade}b shows the stable defect distribution at \qty{140}{\pico\second}, revealing 121 residual point defects, including vacancies and interstitial atoms.
The maximum cluster sizes for vacancies and interstitials are 15 and 11, respectively.
In comparison, a previous study \cite{Liu2023prb} on \emph{elemental} W at similar simulation conditions reported 183 residual point defects with a maximum defect-cluster size exceeding 200 atoms.
The \gls{mpea} thus features fewer defects and smaller defect clusters.
Our simulation results are consistent with the experimental study of a similar tungsten-based refractory \gls{mpea}, which exhibits exceptional radiation resistance, negligible radiation hardening, and no evidence of radiation-induced dislocation loops even at a high dose level \cite{Atwani2019sciadv}.

The enhanced radiation resistance of the tungsten-based refractory \glspl{mpea} could be attributed to the increased chemical complexity, leading to cascade splitting, as depicted in \autoref{fig:cascade}a.
Cascade splitting results in the formation of smaller defect clusters and a more dispersed distribution of isolated (non-clustered) point defects.
This specific application of our \gls{unep1} model to study primary radiation damage through extensive \gls{md} simulations involving 16 million atoms provides further evidence of the generality and high efficiency of our approach.
However, more detailed investigations are necessary to comprehensively characterize and understand the role of alloying in influencing radiation resistance.

\vspace{0.5cm}

\noindent\textbf{Comparisons between \gls{unep1} and \gls{eam} models in \gls{md} and \gls{mcmd} simulations.}
Finally, we showcase the reliability of \gls{unep1} in large-scale \gls{md} and \gls{mcmd} (see Methods) simulations across three applications, providing close comparisons with \gls{eam} results and experimental data.

\begin{figure*}[htb]
\centering
\includegraphics[width=1.5\columnwidth]{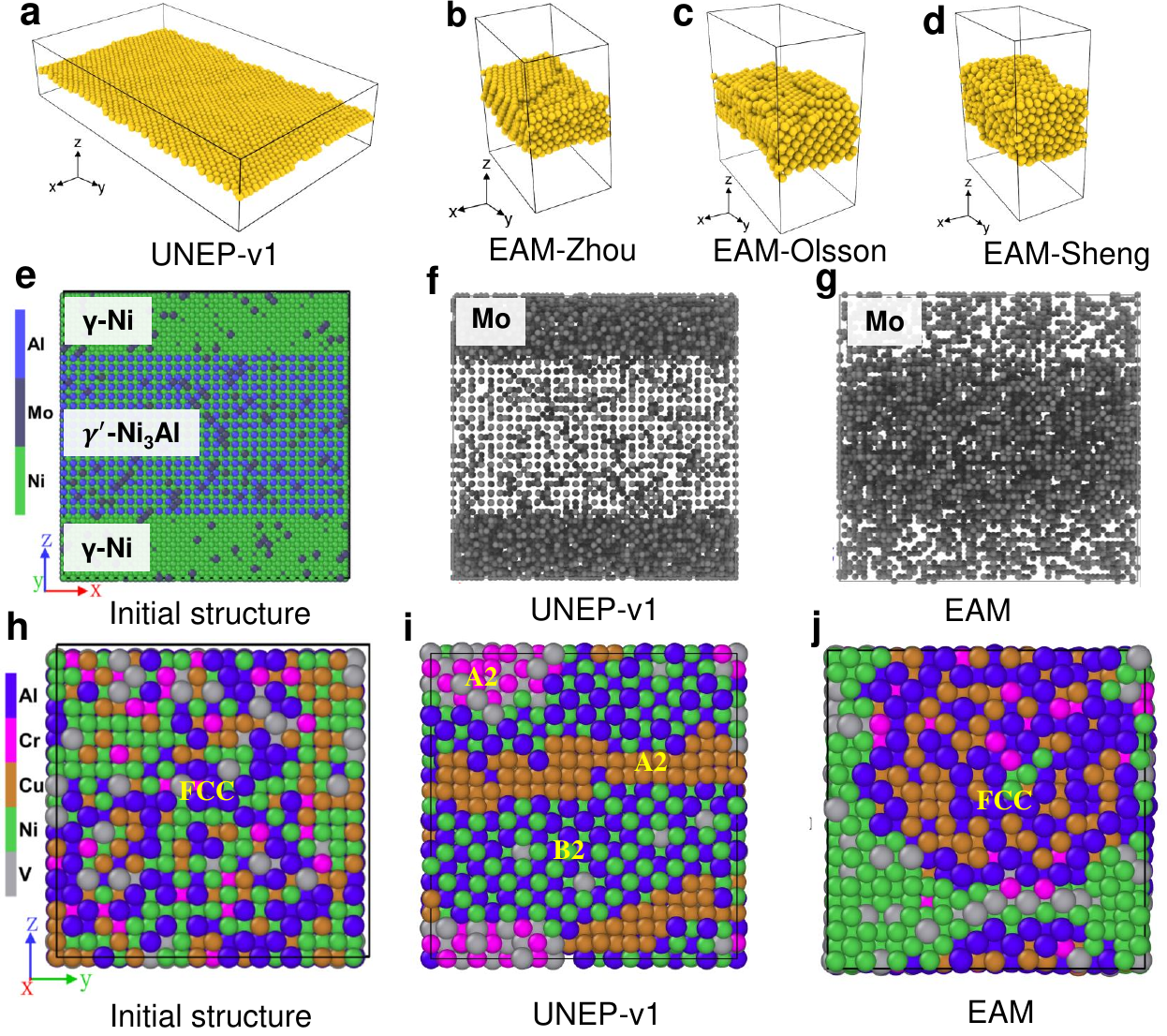}
\caption{
    \textbf{a}--\textbf{d} Snapshots of structures from \gls{md} simulations of \qty{1}{\nano\second} in the isothermal-isobaric ensemble with a target temperature of \qty{300}{\kelvin} and a target in-plane pressure of \qty{0}{\giga\pascal}, starting with a flat monolayer of goldene, using \gls{unep1} and the \gls{eam} potentials by Zhou \textit{et al.} \cite{zhou2004prb}, Olsson  \cite{Olsson2010}, and Sheng \textit{et al.} \cite{Sheng2011prb}. The results demonstrate that \gls{unep1} can maintain goldene's 2D structure at ambient temperature and pressure, while the three \gls{eam} potentials cannot.
    \textbf{e} Initial structure of a $\gamma$-Ni and $\gamma'$-Ni$_3$Al superlattice with a random Mo distribution. 
    \textbf{f}--\textbf{g} Snapshots of the final equilibrium Mo distributions from \gls{mcmd} simulations using \gls{unep1} and \gls{eam} models \cite{zhou2004prb}. \gls{unep1} correctly reproduces the final ratio of the Mo concentration in $\gamma'$-Ni$_3$Al to that in $\gamma$-Ni ($K^{\gamma'/\gamma}=0.667$), in good agreement with experimental observations, while the \gls{eam} potential by Zhou \textit{et al.} \cite{zhou2004prb} gives a value ($K^{\gamma'/\gamma}=4.981$) contradicting experimental trend.
    \textbf{h} Initial \gls{fcc} structure of Al$_{0.31}$Cr$_{0.06}$Cu$_{0.22}$Ni$_{0.32}$V$_{0.09}$.
    \textbf{i}--\textbf{j} Snapshots of the final equilibrium structures from \gls{mcmd} simulations using \gls{unep1} and \gls{eam} models \cite{zhou2004prb}. \gls{unep1} successfully produces both disordered (A2) and ordered (B2) \gls{bcc} structures in full agreement with experiments \cite{YI2020jac}. In contrast, \gls{eam} potential by Zhou \textit{et al.} \cite{zhou2004prb} keeps the system in the \gls{fcc} structure, unable to reproduce the experimentally expected \gls{bcc} structure. (See Fig.~S24 for similar results for Al$_{0.20}$Cr$_{0.12}$Cu$_{0.19}$Ni$_{0.35}$V$_{0.14}$.)
}
\label{fig:UNEPVSEAM}
\end{figure*}

In the first application, we use \gls{unep1} to perform \gls{md} simulations for the recently synthesized goldene \cite{kashiwaya2024synthesis}, a monolayer form of gold that is not explicitly included in the training dataset.
The stable configuration of goldene features a triangular lattice.
We first construct a rectangular cell with \num{1800} atoms and then performed \gls{md} simulations in the isothermal-isobaric ensemble with a target temperature of \qty{300}{\kelvin} and a target in-plane pressure of \qty{0}{\giga\pascal}.
Figure \ref{fig:UNEPVSEAM}a shows that \gls{unep1} maintains the stability of the goldene sheet at \qty{300}{\kelvin}, exhibiting out-of-plane ripples typical for two-dimensional materials.
In contrast, \autoref{fig:UNEPVSEAM}b--d shows that the monolayer structure of goldene cannot be maintained by \gls{eam} potentials from the literature, which include the one used for most benchmarks \cite{zhou2004prb} and two more recent ones \cite{Olsson2010, Sheng2011prb}.
The trajectories for the \gls{md} simulations using the \gls{unep1} model and the three \gls{eam} potentials are presented as movies in the SI.
The results here demonstrate that the \gls{unep1} model has good generalizability in the configuration space.

In the second application, we use \gls{mcmd} simulations to study the Mo distribution in a superlattice structure formed by $\gamma$-Ni and $\gamma'$-Ni$_3$Al. 
Starting from a uniform Mo distribution with an overall Mo concentration of 8.1\% (\autoref{fig:UNEPVSEAM}e), the final ratio of the Mo concentration in $\gamma'$-Ni$_3$Al to that in $\gamma$-Ni is $K^{\gamma'/\gamma}=0.667$ according to our \gls{unep1} model (\autoref{fig:UNEPVSEAM}f).
This agrees well with experimental observations indicating $K^{\gamma'/\gamma}<1$ when the initial Mo concentration is above approximately 6\% \cite{Tu2012apl, jia1994MMTA}.
In contrast, the \gls{eam} potential by Zhou \textit{et al.} \cite{zhou2004prb} gives a value of $K^{\gamma'/\gamma}=4.981$ (\autoref{fig:UNEPVSEAM}g), which contradicts the experimental trend.

In the third application, we use \gls{mcmd} simulations to reproduce the experimentally expected \gls{bcc} structure in the Al-rich intermetallic Al$_{0.31}$Cr$_{0.06}$Cu$_{0.22}$Ni$_{0.32}$V$_{0.09}$, despite the presence of a large fraction of \gls{fcc} metals, starting from an initial \gls{fcc} structure (\autoref{fig:UNEPVSEAM}h). 
Our \gls{unep1} model successfully produces both disordered (A2) and ordered (B2) \gls{bcc} structures (\autoref{fig:UNEPVSEAM}i) in full agreement with experiments \cite{YI2020jac}.
In contrast, the \gls{eam} potential by Zhou \textit{et al.} \cite{zhou2004prb} keeps the system in the \gls{fcc} structure (\autoref{fig:UNEPVSEAM}j). 
Similar results for Al$_{0.20}$Cr$_{0.12}$Cu$_{0.19}$Ni$_{0.35}$V$_{0.14}$ are shown in Fig.~S24, further demonstrating the superior reliability of \gls{unep1} over the \gls{eam} model.
Finally, in Fig.~S25, we illustrate that the equimolar TiZrVMo and TiZrVMoTa alloys correctly transform to \gls{bcc} structures from \gls{hcp} structures during \gls{mcmd} simulations with \gls{unep1}, in excellent agreement with experimental observations \cite{Mu2017@jac}.
This indicates that the \gls{unep1} model, trained on 1-component and 2-component structures, can correctly capture phase transitions occurring in multi-element alloys.

\section{Discussion}

In summary, we have developed an advanced \gls{nep} approach capable of constructing accurate and efficient general-purpose \glspl{mlp} for numerous elements and their alloys.
Two crucial extensions have been made compared to previous \gls{nep} versions.
Firstly, we employed distinct \glspl{nn} for each species, ensuring consistent regression capacity even as the number of species grows.
Secondly, we introduced multiple loss functions to optimize different subsets of the parameters, crucially accelerating the training process when using evolutionary algorithms with a large number of trainable parameters.
We expect that this concept can more generally boost the application of evolutionary algorithms in solving complex optimization problems.

A pivotal insight driving the success of this approach is the recognition that chemical (species) information can be embedded in the trainable expansion coefficients of radial functions, dependent only on atom pairs and basis functions.
As a result, the 1-component and 2-component structures delineate an outer boundary in descriptor space, while $n$-component structures with $n \geq 3$ represent interpolation points in this space.
Leveraging the exceptional interpolation capabilities of \glspl{nn}, a \gls{nep} model trained solely with 1-component and 2-component structures performs very well for $n$-component structures with $n \geq 3$, provided the configuration space has been sufficiently explored.
The effectiveness of this approach has been demonstrated through accurate predictions of formation energies across various multi-component alloys, as well as  reproduction of experimentally observed chemical order and stable phases, using our \gls{unep1} model.

While the current study focuses on 16 elements, our approach is scalable and adaptable for constructing \gls{nep} models across the entire periodic table.
The primary challenge resides in the generation of the reference data, typically via \gls{dft} calculations, rather than the regression capabilities of the \gls{nep} model.
Notably, our approach is also sustainable.
Starting from our existing training set for 16 elements, one merely needs to include structures involving 17 chemical compositions (one 1-component and 16 2-component systems) to form a comprehensive training set for 17 elements.
This method is far more economical than building an entirely new training set from scratch.
Beyond extending the chemical space, one can also broaden the configuration space for existing chemical compositions, through established active-learning approaches, especially with the aid of structure searching methods \cite{wang2023magus}. 

The successful applications of the \gls{unep1} model in studying plasticity and primary radiation damage in the MoTaVW refractory \glspl{mpea} demonstrate the versatility and robustness of the NEP4 approach in general and the \gls{unep1} model in particular, establishing its significant potential for in-depth explorations and insights into the intricate behavior of complex materials such as \glspl{mpea}.

In conclusion, our study demonstrates the promise of our approach in constructing a unified general-purpose \gls{mlp} for the periodic table with remarkable computational efficiency, taking full advantage of the embedded chemical generalizability, the outstanding interpolation capabilities of \glspl{nn} and an advanced multiple-loss evolutionary training algorithm for many-component systems.
By successfully developing a highly accurate and efficient \gls{mlp} for a diverse range of elemental metals and alloys, our study showcases the versatility and applicability of our approach across various materials.
These advancements mark a significant leap forward in enhancing the practical applications of \glspl{mlp} in materials modeling, offering new opportunities for more accurate, efficient, and predictive computer simulations in materials research.

\section{Methods}

\noindent{\textbf{MD simulations for training structure generation.}}
To create the initial training structures, we used the \textsc{lammps} package (23 Jun 2022) \cite{thompson2022cpc} to run \gls{md} simulations with cells ranging from 32 to 108 atoms. 
For each 1-component or 2-component system, we ran \gls{md} simulations in the isothermal-isobaric ensemble (zero target pressure) using the \gls{eam} potential \cite{zhou2004prb} at 9 temperatures (50, 300, 800, \num{1300}, \num{1700}, \num{2300}, \num{3000}, \num{4000}, and \qty{5000}{\kelvin}), each for \qty{2}{\nano\second}.
For each \gls{md} run, we sampled 5 structures. 
For each structure, we made three copies, one with a subsequent box scaling of 95\%, one with 105\%, and one with 5\% (random) box perturbation. 
We also ran \gls{md} simulations at \qty{300}{\kelvin} with  tensile or compressing loading with a strain rate of \qty{2e8}{\per\second} for 1 to \qty{2}{\nano\second} and uniformly sampled 35 structures.

\vspace{0.5cm}

\noindent{\textbf{DFT calculations for reference data generation.}}
After preparing the initial training structures, we performed quantum-mechanical calculations to obtain reference data, including the energy and virial for each structure and the force on each atom in each structure. 
\Gls{dft} calculations as implemented in \textsc{vasp} \cite{Kresse1996prb} were performed to generate reference data.
The \verb"INCAR" file for \textsc{vasp} is presented in Supplementary Note S1.

We used the projector augmented wave method \cite{paw1, paw2}, the PBE functional \cite{Perdew1996prl}, an energy cutoff of \qty{600}{\electronvolt} , a $\Gamma$-centered $k$-point mesh with a spacing of \qty{0.2}{\per\angstrom}, and a threshold of \qty{e-6}{\electronvolt} for the electronic self-consistent loop.
We used the blocked Davidson iteration scheme for electronic minimization.
The \verb"PREC" tag in the \textsc{vasp} input file was set to \verb"Accurate" to ensure accurate forces.
A Gaussian smearing with a smearing width of \qty{0.02}{\electronvolt} was used.
The Gaussian smearing is not the best choice for elemental metals and their alloys but we chose this in view of possible future extension of our approach to the whole periodic table.
Our settings can ensure a convergence of the energy to \qty{1}{\milli\electronvolt\per\atom} for all the materials. In our DFT calculations, we did not consider magnetism, consistent with previous works \cite{zhao2023md,Lopanitsyna2023PRM}. We have tested that modeling Ni and Cr as ferromagnetic does not change the energy ordering for the major phases, see Fig. S26. While a proper account of magnetism might help to improve the quality of the DFT results and the resulting potential, it would significantly complicate the training database and require extensive additional computational resources. 

\vspace{0.5cm}

\noindent{\textbf{The NEP training hyperparameters.}}
We used \textsc{gpumd} v3.9.3 to train the \gls{unep1} model, which is a NEP4 model as introduced in this paper.
The details of the \verb"nep.in" input file we used and the \gls{snes}-based multi-loss training algorithm can be found in Supplementary Note S2.

The cutoff radii for radial and angular descriptor parts are \qty{6}{\angstrom} and \qty{5}{\angstrom}, respectively.
For both the radial and the angular descriptor components, we used 5 radial functions constructed from a linear combination of 9 basis functions. 
The descriptor vector for one element thus has $5 + 5 \times 6 = 35$ components.
There are 80 neurons used in the hidden layer and the \gls{nn} architecture for each element can be written as $35$-$80$-$1$, corresponding to \num{2960} trainable parameters.
For each pair of elements, there are $5\times 9+5\times 9=90$ trainable descriptor parameters.
The total number of trainable parameters in the \gls{unep1} model for 16 elements is thus $2960 \times 16 + 90 \times 16^2 + 1 = \num{70401}$, where a global bias (shifting) parameter is included.
The training was performed with a batch size of \num{10000} structures for \num{1000000} generations (steps), which took about ten days using four A100 GPUs.

\vspace{0.5cm}

\noindent{\textbf{Calculations of basic physical properties.}}
To evaluate the reliability of the \gls{unep1} model in molecular statics and \gls{md} simulations, we calculated a set of relevant static and dynamic material properties, with a close comparison with \gls{eam} \cite{zhou2004prb}, \gls{dft} (if available or affordable) \cite{De2015,Tran2016surface}, and experiments (if available).
Energetics, elastic properties, and phonon dispersion relations were calculated with the help of \textsc{gpumd-wizard}, \textsc{ase} \cite{Hjorth_Larsen_2017}, \textsc{pynep} \cite{fan2022jcp}, \textsc{calorine} \cite{Lindgren2024joss}, and \textsc{phonopy} \cite{Togo2023JPCM} packages.
Melting points were calculated using the two-phase method as implemented in \textsc{gpumd} \cite{fan2022jcp} for \gls{unep1} and \textsc{lammps} \cite{thompson2022cpc} for \gls{eam}, and are compared to experimental values \cite{CRChandbook}.
Vacancy formation energies were evaluated using $4 \times 5 \times 6$ supercells.
The formation energies of free surfaces were evaluated with $2 \times 2 \times 10$ supercells (taking a surface perpendicular to $z$ as an example here). 
The uncertainties in the predictions for the bulk and shear moduli and volume per atom, over different ensemble models, were estimated as the standard deviation using \textsc{calorine} \cite{Lindgren2024joss}.

\vspace{0.5cm}

\noindent{\textbf{MD simulations for plasticity of MPEAs.}}
We used the \gls{unep1} model to drive \gls{md} calculations of the plasticity of \glspl{mpea} under compression using the \textsc{gpumd} package \cite{fan2022jcp, fan2017cpc}.
First, we used the Voronoi algorithm implemented in \textsc{atomsk} \cite{atomsk} to build our initial MoTaVW polycrystalline \gls{mpea} model by removing overlapping atoms at boundaries.
The model is composed of 12 grains with sizes ranging from \qty{96}{\nano\meter\cubed} to \qty{195}{\nano\meter\cubed}, and contains \num{100 205 176} atoms which randomly occupy a \gls{bcc} lattice at equimolar ratios.
The initial MoTaVW model was further relaxed by \gls{md} simulations for \qty{500}{\pico\second} in the isothermal-isobaric ensemble at \qty{300}{\kelvin} and \qty{0}{\giga\pascal} using the Bernetti-Bussi barostat \cite{Bernetti2020jcp} and Bussi-Donadio-Parrinello thermostat \cite{Bussi2007jcp}.
Finally we simulated uniaxial compressive deformation with a constant engineering strain rate of \qty{4.2e8}{\per\second}.
The time step was kept fixed at \qty{1}{\femto\second}.
The 2D visualization of dislocations perpendicular to the compressive axis was rendered using the \textsc{ovito} package \cite{ovito}.

\vspace{0.5cm}

\noindent{\textbf{MD simulations for primary radiation damage.}}
The \gls{md} simulations of the displacement cascade in MoTaVW were performed using the \textsc{gpumd} package \cite{fan2022jcp, fan2017cpc} with the \gls{unep1} model and a repulsive two-body \gls{zbl}-like potential \cite{Ziegler1985}.
A periodic cubic simulation cell with \num{16000000} atoms was constructed by creating a random mixture of the Mo, Ta, V, and W atoms with equimolar ratio in a \gls{bcc} crystal.
We equilibrated this system in the isothermal-isobaric ensemble for \qty{30}{\pico\second}, with a target temperature of \qty{300}{\kelvin} and a target pressure of \qty{0}{\giga\pascal}.
A primary knock-on atom with an energy of \qty{100}{\kilo\electronvolt} moving in the high-index direction $\langle 135 \rangle$ (to avoid channeling effects) was then created at the center of the simulation cell.
Atoms within a thickness of three lattice constants of the boundaries were maintained at \qty{300}{\kelvin}.
The integration time step had an upper limit of \qty{1}{\femto\second} and was dynamically determined so that the fastest atom could move at most \qty{0.015}{\angstrom} (less than 0.5\% of the lattice constant) within one step.
The total number of steps is \num{200000}, corresponding to \qty{140}{\pico\second}.
Electronic stopping \cite{Kai1995cms} was applied as a frictional force on atoms with a kinetic energy over \qty{10}{\electronvolt}.
We used the \textsc{ovito} package \cite{ovito} for defect analysis and visualization.
The interstitials and vacancies were identified by using the Wigner-Seitz cell method.
The defects were grouped into clusters: two vacancies were considered to be in the same cluster if the distance between them was within the second-nearest-neighbor distance, while the third-nearest-neighbor distance was used to identify self-interstitial clusters.

\vspace{0.5cm}

\noindent{\textbf{MCMD simulations for multi-component alloys.}}
We utilized \gls{mcmd} simulations in the canonical \gls{mc} ensemble (involving swapping atoms of different species) \cite{song2024solute} in the study of Mo distribution in a superlattice structure formed by $\gamma$-Ni and $\gamma'$-Ni$_3$Al (\num{108000}-atom supercell), the \gls{fcc}-to-\gls{bcc} transformation in Al-rich alloys (Al$_{0.31}$Cr$_{0.06}$Cu$_{0.22}$Ni$_{0.32}$V$_{0.09}$ and Al$_{0.20}$Cr$_{0.12}$Cu$_{0.19}$Ni$_{0.35}$V$_{0.14}$, with \num{4000}-atom supercell), and the \gls{hcp}-to-\gls{bcc} transformation in the equimolar TiZrVMo and TiZrVMoTa alloys (\num{1600}-atom supercell). During these \gls{mcmd} simulations, \gls{mc} trials are attempted 2000 times after every 1000 MD steps, totaling about $10^6$ \gls{md} steps to reach equilibrium, at which the \gls{mc} acceptance ratio is close to zero. 

\vspace{0.5cm}

\section*{Data availability}

The training data and trained NEP models are freely available at the Zenodo repository \url{https://doi.org/10.5281/zenodo.10081676}. 
The high-throughput-calculation inputs/outputs for the basic physical properties are freely available at \url{https://github.com/Jonsnow-willow/GPUMD-Wizard}. 

\section*{Code availability}

The source code and documentation for \textsc{calorine} are available at \url{https://gitlab.com/materials-modeling/calorine} and \url{https://calorine.materialsmodeling.org/}, respectively.
The source code and documentation for \textsc{gpumd} are available
at \url{https://github.com/brucefan1983/GPUMD} and \url{https://gpumd.org}, respectively. 
The source code and documentation for \textsc{gpumd-wizard} are available at \url{https://github.com/Jonsnow-willow/GPUMD-Wizard}.
The source code and documentation for \textsc{pynep} are available at \url{https://github.com/bigd4/PyNEP} and \url{https://pynep.readthedocs.io/en/latest/}, respectively. 

\section*{Declaration of competing interest}
The authors declare that they have no competing interests.

\section*{Contributions}

Keke Song, Rui Zhao, Yanzhou Wang, and Nan Xu prepared the training and test structures. 
Keke Song, Yanzhou Wang, Zhiqiang Zhao, Ting Liang, Jiuyang Shi, Junjie Wang, Ke Xu, Shuang Lyu, Zezhu Zeng, and Shirong Liang performed the DFT calculations.  
Keke Song, Shunda Chen, and Yanzhou Wang tested the various hyperparameters and trained the NEP models. 
Penghua Ying analyzed the descriptor space.
Keke Song, Rui Zhao, Jiahui Liu, Ke Xu, Ting Liang, Zhiqiang Zhao, and Haikuan Dong evaluated the basic physical properties.
Eric Lindgren trained the first generation of ensemble models, and performed the ensemble model analysis. 
Yong Wang and Shunda Chen evaluated the computational performance. 
Jiahui Liu performed the radiation damage simulations and developed the high-throughput calculation tools. 
Zheyong Fan and Shunda Chen developed the NEP4 model and the multiple-loss evolutionary training algorithm. 
Ligang Sun, Yue Chen, Zhuhua Zhang, Wanlin Guo, Ping Qian, Jian Sun, Paul Erhart, Tapio Ala-Nissila, Yanjing Su, and Zheyong Fan supervised the project.

\section*{Acknowledgements}
Keke Song, Jiahui Liu, Yanzhou Wang, Ping Qian, and Yanjing Su acknowledge support from the National Key R \& D Program of China (No. 2022YFB3707500) and the National Natural Science Foundation of China (NSFC) (No. 92270001).
Yanzhou Wang and Tapio Ala-Nissila have been supported in part by the Academy of Finland through its Quantum Technology Finland CoE grant No. 312298 and under the European Union – NextGenerationEU instrument by the Academy of Finland grant 353298.
Eric Lindgren and Paul Erhart acknowledge funding from the Swedish Research Council (Nos.~2020-04935 and 2021-05072) and the Swedish Foundation for Strategic Research via the SwedNESS graduate school (GSn15-0008) as well as computational resources provided by the National Academic Infrastructure for Supercomputing in Sweden at NSC and C3SE partially funded by the Swedish Research Council through grant agreement No.~2022-06725.
Jian Sun acknowledges support from NSFC (Nos. 12125404, 11974162), the Basic Research Program of Jiangsu, the Fundamental Research Funds for the Central Universities, and computational sources from the High Performance Computing Center of Collaborative Innovation Center of Advanced Microstructures and the high-performance supercomputing center of Nanjing University.
Ke Xu and Ting Liang acknowledge support from the National Key R\&D Project from Ministry of Science and Technology of China (No. 2022YFA1203100), the Research Grants Council of Hong Kong (No. AoE/P-701/20), and RGC GRF (No. 14220022). 
Zhiqiang Zhao, Zhuhua Zhang, and Wanlin Guo acknowledge support from the NSFC Projects of International Cooperation and Exchanges (No. 12261160367).
Zezhu Zeng, Shuang Lyu, and Yue Chen are grateful for the research computing facilities offered by ITS, HKU. 

\bibliography{refs}

\clearpage
\onecolumngrid

\section{Supplementary Notes}

\renewcommand{\refname}{}
\renewcommand{\figurename}{Figure}
\renewcommand{\tablename}{Table}
\renewcommand\thefigure{S\arabic{figure}}
\renewcommand\thetable{S\arabic{table}}
\renewcommand\thenote{S\arabic{note}}
\renewcommand\theequation{S\arabic{equation}}

\renewcommand{\topfraction}{.9}
\renewcommand{\bottomfraction}{.9}
\renewcommand{\floatpagefraction}{.9}

\renewcommand{\listfigurename}{Supplementary Figures}
\renewcommand{\listtablename}{Supplementary Tables}

\setcounter{figure}{0}
\setcounter{table}{0}

\notetitlelabel{snote:incar}{The INCAR input file for VASP}

We have used the following inputs in the \verb"INCAR" file of the \textsc{vasp} code for training data calculations.

\begin{verbatim}
  GGA      = PE       # Default is POTCAR
  ENCUT    = 600      # Default is POTCAR
  KSPACING = 0.2      # Default is 0.5
  KGAMMA   = .TRUE.   # This is the default
  NELM     = 120      # Default is 60
  ALGO     = Normal   # This is the default
  EDIFF    = 1E-06    # Default is 1E-4
  SIGMA    = 0.02     # Default is 0.2
  ISMEAR   = 0        # Default is 1
  PREC     = Accurate # Default is Normal
  LREAL    = A        # Default is .FALSE.
\end{verbatim}

\notetitlelabel{snote:hyper}{The nep.in input file for GPUMD and details of the multi-loss SNES training algorithm}

We have used the following inputs in the \verb"nep.in" file of the \textsc{gpumd} code to train UNEP-v1:

\begin{verbatim} 
type       16 Ag Al Au Cr Cu Mg Mo Ni Pb Pd Pt Ta Ti V W Zr
version    4
cutoff     6 5
n_max      4 4
basis_size 8 8
l_max      4 2 1
neuron     80
lambda_1   0
lambda_e   1
lambda_f   1
lambda_v   0.1
batch      10000
population 60
generation 1000000
zbl        2
\end{verbatim}

We explain the hyperparameters one by one:
\begin{itemize}
\item The \verb|type| keyword specifies the number and chemical symbols of the species we considered.
\item The \verb|version| keyword specifies the version of the NEP model used, which is NEP4 introduced in the present work.
\item The \verb|cutoff| keyword specifies that the cutoff radii for the radial (2-body) and angular (from 3-body to 5-body) descriptor components are 6 and 5 \AA{} respectively.
\item The \verb|n_max| keyword specifies that the number of radial functions for the radial and angular descriptor components are both $4+1=5$.
\item The \verb|basis_size| keyword specifies that the number of radial basis functions for the radial and angular descriptor components are both $8+1=9$.
\item The \verb|l_max| keyword specifies that the levels of the spherical harmonics for the 3-body, 4-body, and 5-body angular descriptor components are up to 4, 2, and 1 respectively.
\item The \verb|neuron| keyword specifies that the number of neurons in the hidden layer is 80.
\item The \verb|lambda_1| keyword specifies that no $\mathcal{L}_1$ regularization is applied here, but the $\mathcal{L}_2$ regularization is applied by default.
\item The \verb|lambda_e| keyword specifies the weight of the energy loss term to be 1.
\item The \verb|lambda_f| keyword specifies the weight of the force loss term to be 1.
\item The \verb|lambda_v| keyword specifies the weight of the virial loss term to be 0.1.
\item The \verb|batch| keyword specifies the batch size of training to be \num{10000}.
\item The \verb|population| keyword specifies the population size to be 60.
\item The \verb|generation| keyword specifies the total number of training generations (steps) to be \num{1000000}.
\item The \verb|zbl| keyword specifies to use the NEP-ZBL model \cite{Liu2023prb} with a cutoff radius of 2 \AA{} for the universal ZBL potential.
\end{itemize}

The UNEP-v1 models were trained based on the following loss function:
\begin{equation}
L(\mathbf{z}) = L _{\rm e}(\mathbf{z}) + L _{\rm f}(\mathbf{z}) + L _{\rm v}(\mathbf{z}) + L _2(\mathbf{z}),
\end{equation}
\begin{equation}
L _{\rm e}(\mathbf{z}) 
= \lambda _\mathrm{e} 
\left( \frac{1}{N _\mathrm{str}}\sum _{n=1} ^{N _\mathrm{str}} \left( U^\mathrm{NEP}(n,\mathbf{z}) - U^\mathrm{ref}(n)\right)^2
\right) ^{1/2},
\end{equation}
\begin{equation}
L _{\rm f}(\mathbf{z}) = \lambda _\mathrm{f} \left( \frac{1}{3N} \sum _{i=1} ^{N} \left( \mathbf{F} _i^\mathrm{NEP}(\mathbf{z}) - \mathbf{F} _i^\mathrm{ref}\right)^2 \right) ^{1/2}, 
\end{equation}
\begin{equation}
L _{\rm v}(\mathbf{z}) = \lambda _\mathrm{v} \left( 
   \frac{1}{6N _\mathrm{str}}
   \sum _{n=1} ^{N _\mathrm{str}} \sum _{\mu\nu} \left( W _{\mu\nu}^\mathrm{NEP}(n,\mathbf{z}) - W _{\mu\nu}^\mathrm{ref}(n)\right)^2
   \right) ^{1/2}, 
\end{equation}
\begin{equation}
L _2(\mathbf{z}) = \lambda _2 \left(\frac{1}{N _\mathrm{par}} \sum _{n=1} ^{N _\mathrm{par}} z _n^2\right) ^{1/2}.
\end{equation}
Here, $\mathbf{z}$ denotes the vector formed by the $N_{\rm par}$ trainable parameters in the model, $N_\mathrm{str}$ is the number of structures containing the given species in one batch, $N$ is the total number of atoms in these structures, $U^\mathrm{NEP}(n,\mathbf{z})$ and $W_{\mu\nu}^\mathrm{NEP}(n,\mathbf{z})$ are the energy and virial for the $n$th structure calculated by the current UNEP-v1 models, $U^\mathrm{ref}(n)$ 
and $W_{\mu\nu}^\mathrm{ref}(n)$ are the corresponding reference values, $\mathbf{F}_i^\mathrm{NEP}(\mathbf{z})$ is the force on atom $i$ calculated by the current UNEP-v1 models, 
$\mathbf{F}_i^\mathrm{ref}$ is the corresponding reference value, and $L_2(\mathbf{z})$ is the $\mathcal{L}_2$ regularization term. 
The SNES training algorithm \cite{Schaul2011} with the extension of the multi-loss approach we proposed is given below.
(1) Create a search distribution in the solution space, including a set of mean values $\mathbf{m}$ and variances $\mathbf{s}$. The elements in $\mathbf{m}$ are uniformly distributed from $-1$ to $1$ and those in $\mathbf{s}$ take a constant value of 0.1.
(2) Starting from the above initial distribution, we perform $N_{\rm gen}$ iterations. First create $N_{\rm pop}$ solutions $\mathbf{z}_k$ ($1\leq k \leq N_{\rm pop}$) based on the current $\mathbf{m}$ and $\mathbf{s}$ vectors: 
\begin{equation}
    \mathbf{z}_k \leftarrow \mathbf{m} + \mathbf{s} \odot \mathbf{r}_k.
\end{equation}
Here, $N_{\rm pop}$ is the population size and $\mathbf{r}_k$ is a set of normal-distribution random numbers with zero mean and unit variance. 
Then evaluate the loss values $L(\mathbf{z}_k)$ for all the solutions $\mathbf{z}_k$ in the population and sort the solutions based on the loss values. Here, \textit{each species has its own set of loss values and sorting scheme}, which is the key point in our multi-loss approach.
Last, calculate the natural gradients (where $u_k$ is a set of rank-based utility values \cite{Schaul2011}),
\begin{equation}
        \nabla_{\mathbf{m}} J \leftarrow \sum_{k=1}^{N_{\rm pop}} u_k \mathbf{r}_k,
\end{equation}
\begin{equation}
        \nabla_{\mathbf{s}} J \leftarrow \sum_{k=1}^{N_{\rm pop}} u_k (\mathbf{r}_k \odot \mathbf{r}_k-1),
\end{equation}
and update the means and variances of the search distribution,
\begin{equation}
        \mathbf{m} \leftarrow \mathbf{m} + \eta_{\mathbf{m}} \left(\mathbf{s} \odot \nabla_{\mathbf{m}} J\right),
\end{equation}
\begin{equation}
        \mathbf{s} \leftarrow \mathbf{s} \odot \exp\left(\frac{\eta_{\mathbf{s}}}{2} \nabla_{\mathbf{s}} J \right),
\end{equation}
where $\eta_{\mathbf{m}}=1$, and $\eta_{\mathbf{s}}=
        \left(3+\ln N_{\rm par}\right)/5\sqrt{N_{\rm par}}$.
        
\newpage 

\section{\listfigurename}

\begin{figure}[b]
\centering
\includegraphics[width=\columnwidth]{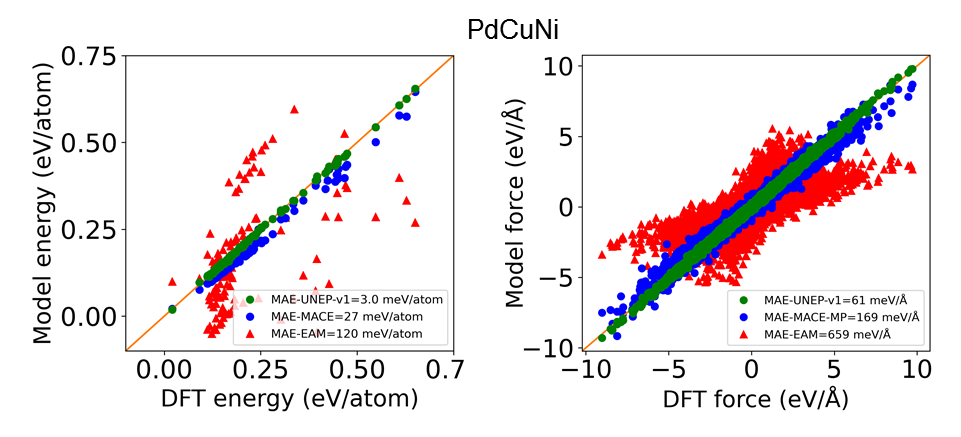}
\caption{
Parity plot for formation energy and 
 force predictions from UNEP-v1, MACE-MP-0 (medium model) and EAM compared to DFT for the test dataset containing up to 3 components (Pd, Cu, Ni) from Zhao \textit{et al.}~\cite{zhao2023md}. UNEP-v1 demonstrates significantly better predictions.
}
\label{fig:zhao2023md}
\end{figure}

\begin{figure}[b]
\centering
\includegraphics[width=\columnwidth]{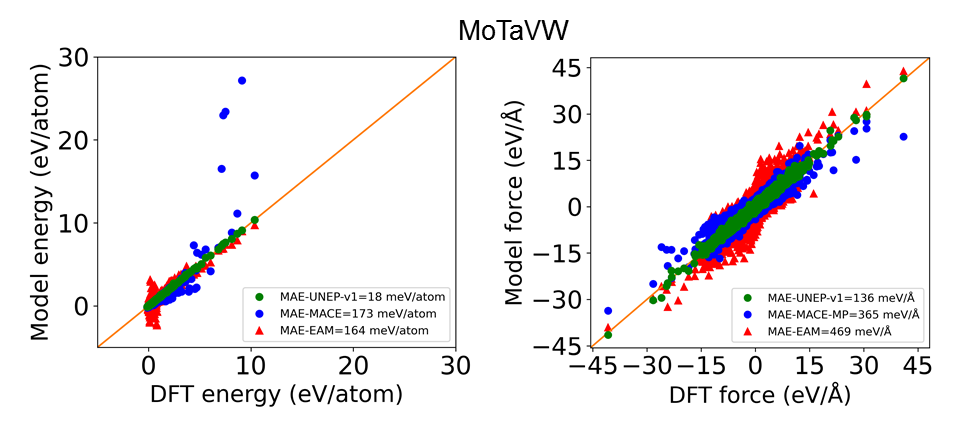}
\caption{
Parity plot for formation energy and force predictions from UNEP-v1, MACE-MP-0 (medium model) and
EAM compared to DFT for the test dataset containing up to 4 components (Mo, Ta, V, W) from Byggm\"astar \textit{et al.} \cite{Byggmastar2022prm}. UNEP-v1 shows much better predictions.
}
\label{fig:Byggmastar2022prm}
\end{figure}

\begin{figure}[b]
\centering
\includegraphics[width=\columnwidth]{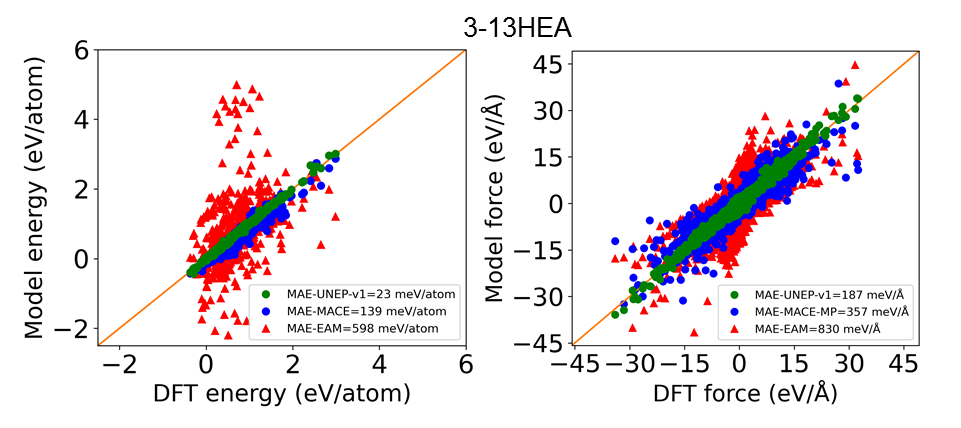}
\caption{
Parity plot for formation energy and force predictions from UNEP-v1, MACE-MP-0 (medium model) and
EAM compared to DFT for the test dataset containing up to 13 components (Ag, Au, Cr, Cu, Mo, Ni, Pd, Pt, Ta, Ti, V, W, Zr) from Lopanitsyna \textit{et al.} \cite{Lopanitsyna2023PRM}. UNEP-v1 demonstrates significantly better predictions.
}
\label{fig:Lopanitsyna2023PRM}
\end{figure}

\begin{figure}[b]
\centering
\includegraphics[width=\columnwidth]{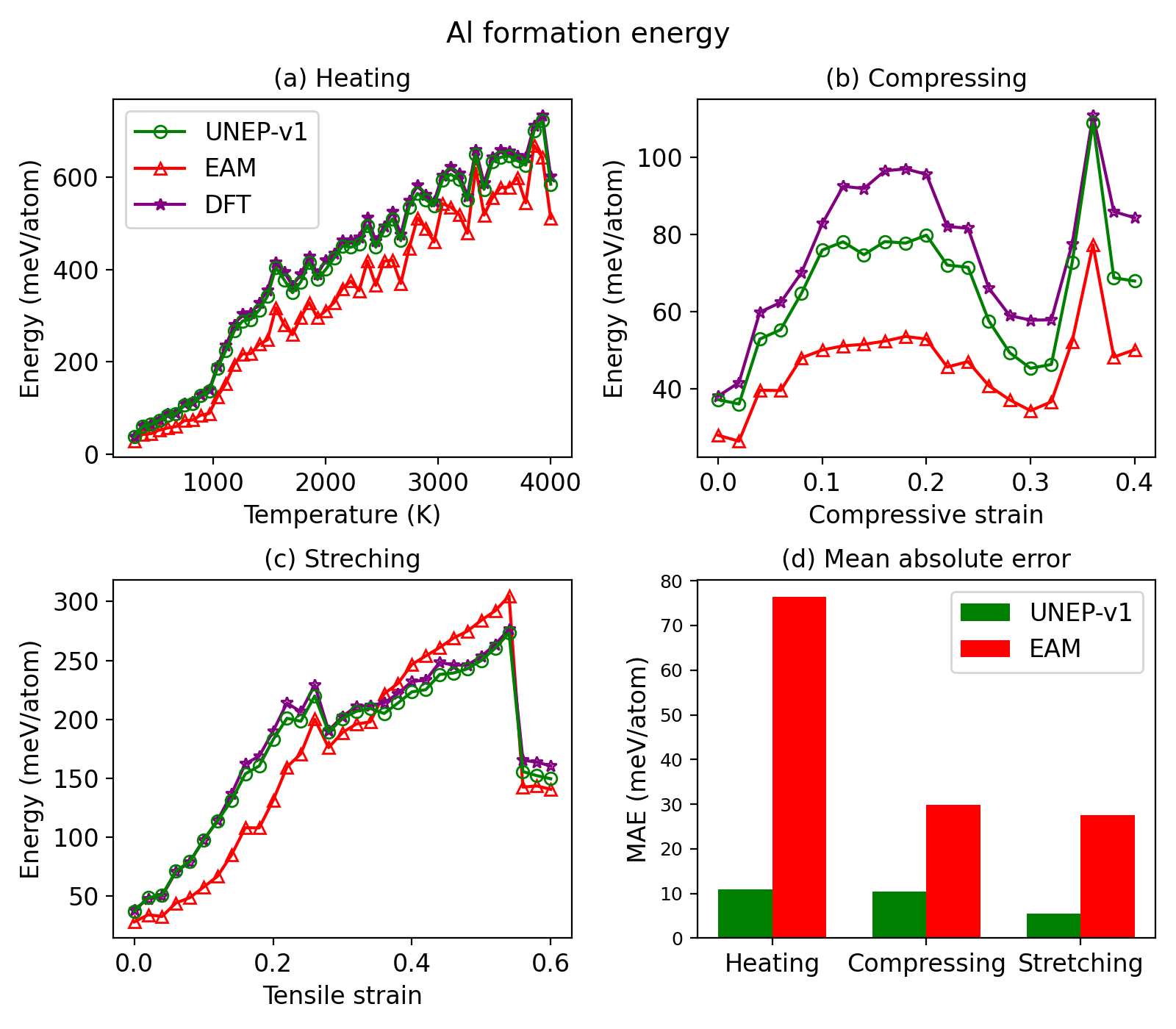}
\caption{
Comparison of the formation energies of FCC Al during (a) a heating process from 300 to 4000 K, (b) a compressing process up to 40\% compressive strain, and (c) a stretching process up to 60\% tensile strain for UNEP-v1, EAM, and DFT. (d) The mean absolute errors (MAEs) for UNEP-v1 and EAM as compared to DFT. The heating rate in (a) is 7.4 K/ps. The deformation rate in (b) and (c) is \num{2e8} s$^{-1}$. The trajectories were sampled using the UNEP-v1 model. UNEP-v1 shows much better predictions than EAM.
}
\label{fig:pure_Al}
\end{figure}

\begin{figure}[b]
\centering
\includegraphics[width=\columnwidth]{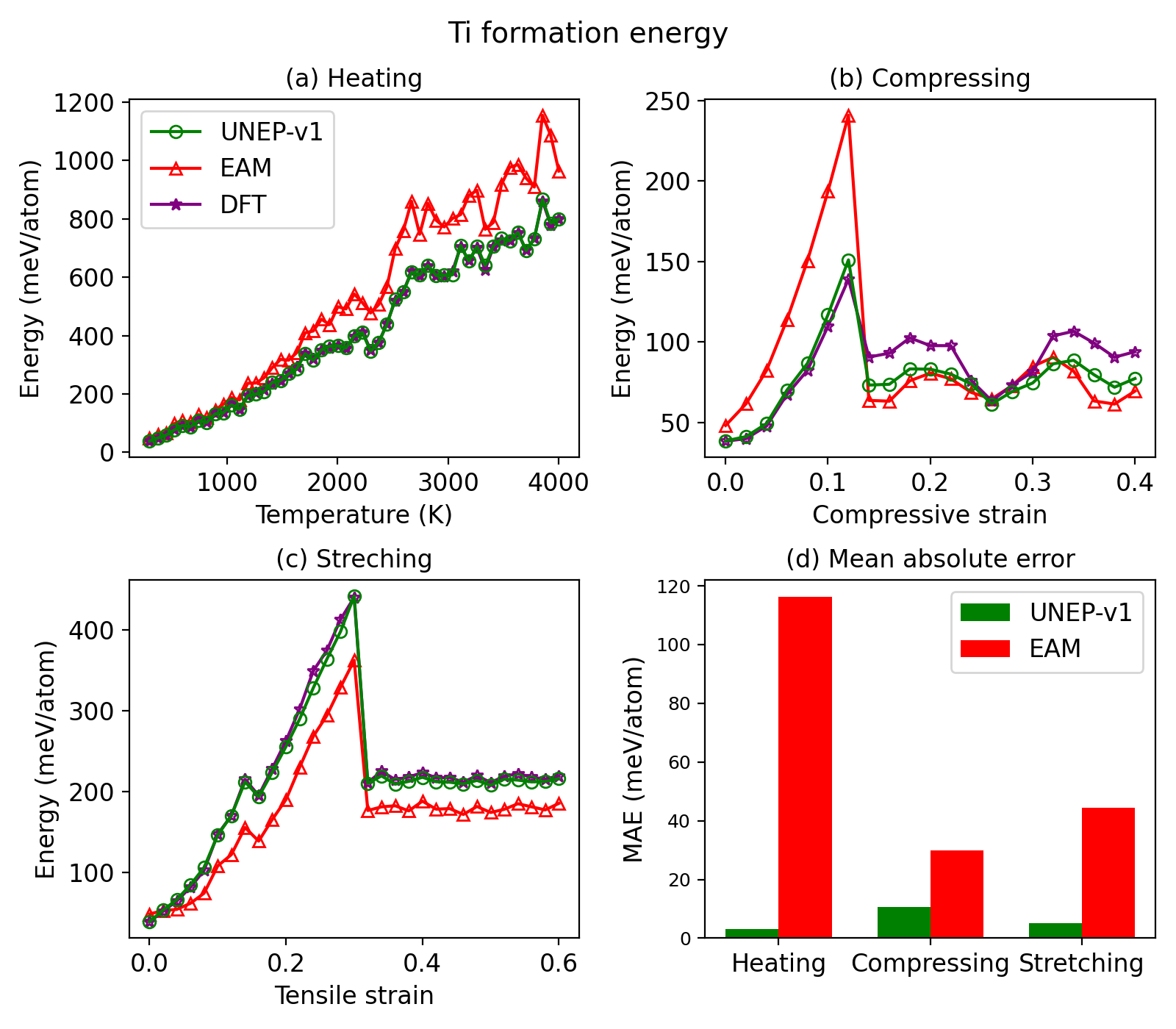}
\caption{
Similar to Fig.~\ref{fig:pure_Al} but for HCP Ti.
}
\label{fig:pure_Ti}
\end{figure}

\begin{figure}[b]
\centering
\includegraphics[width=\columnwidth]{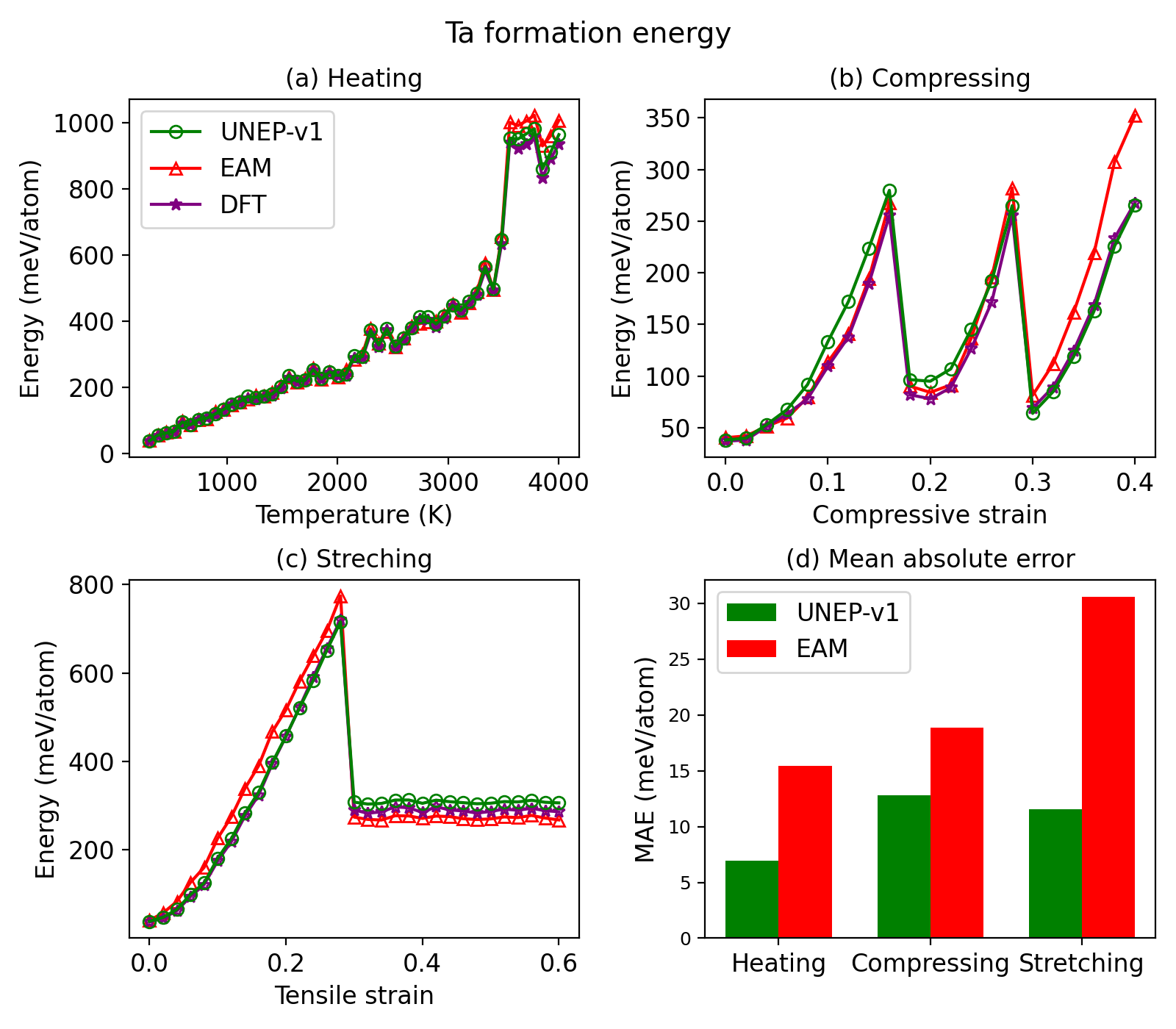}
\caption{
Similar to Fig.~\ref{fig:pure_Al} but for BCC Ta.
}
\label{fig:pure_Ta}
\end{figure}

\begin{figure}[b]
\centering
\includegraphics[width=\columnwidth]{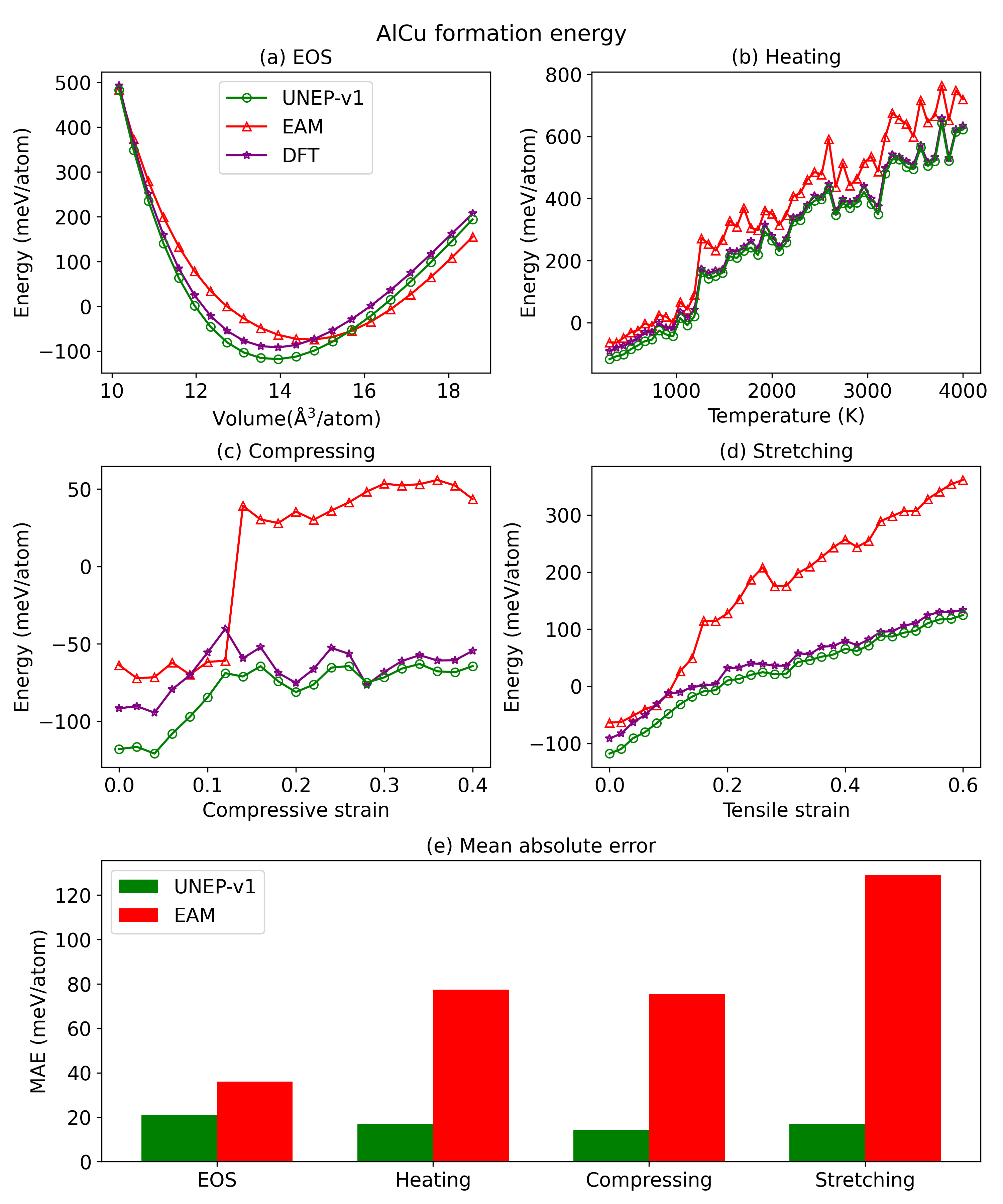}
\caption{
Comparison of the formation energies of FCC AlCu alloy for (a) the equation of state curve, (b) a heating process from 300 to 4000 K, (c) a compressing process up to 40\% compressive strain, and (d) a stretching process up to \%60 tensile strain as calculated from UNEP-v1, EAM, and DFT. (e) The mean absolute errors (MAEs) for UNEP-v1 and EAM as compared to DFT. The heating rate in (a) is 7.4 K/ps. The deformation rate in (b) and (c) is \num{2e8} s$^{-1}$. The trajectories were sampled by using the UNEP-v1 model, with the initial structure chemically relaxed through hybrid Monte-Carlo and molecular dynamics (MCMD) simulations.
}
\label{fig:AlCu_results}
\end{figure}

\begin{figure}[b]
\centering
\includegraphics[width=\columnwidth]{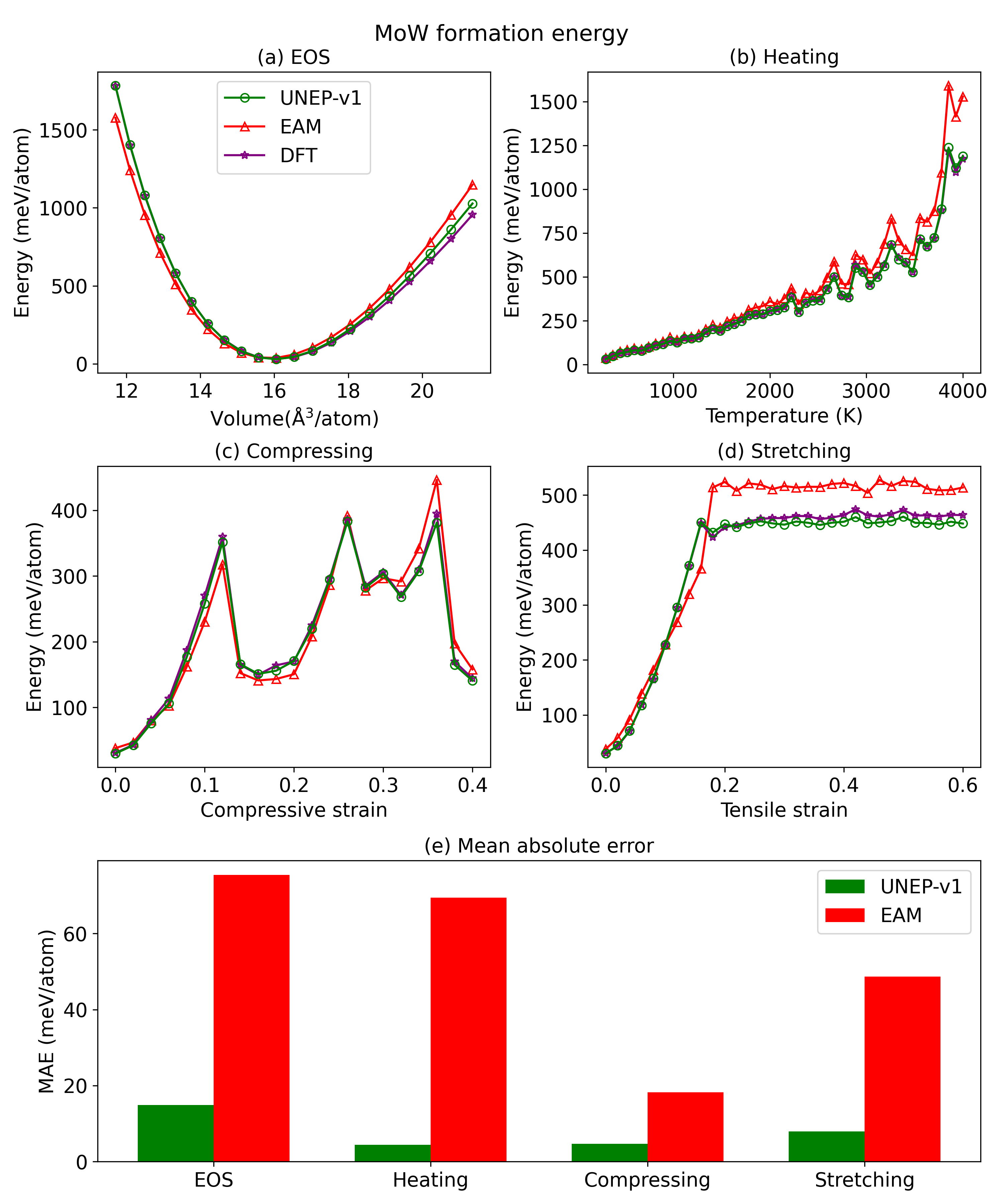}
\caption{
Similar to Fig.~\ref{fig:AlCu_results} but for BCC MoW.
}
\label{fig:WMo_results}
\end{figure}

\begin{figure}[b]
\centering
\includegraphics[width=\columnwidth]{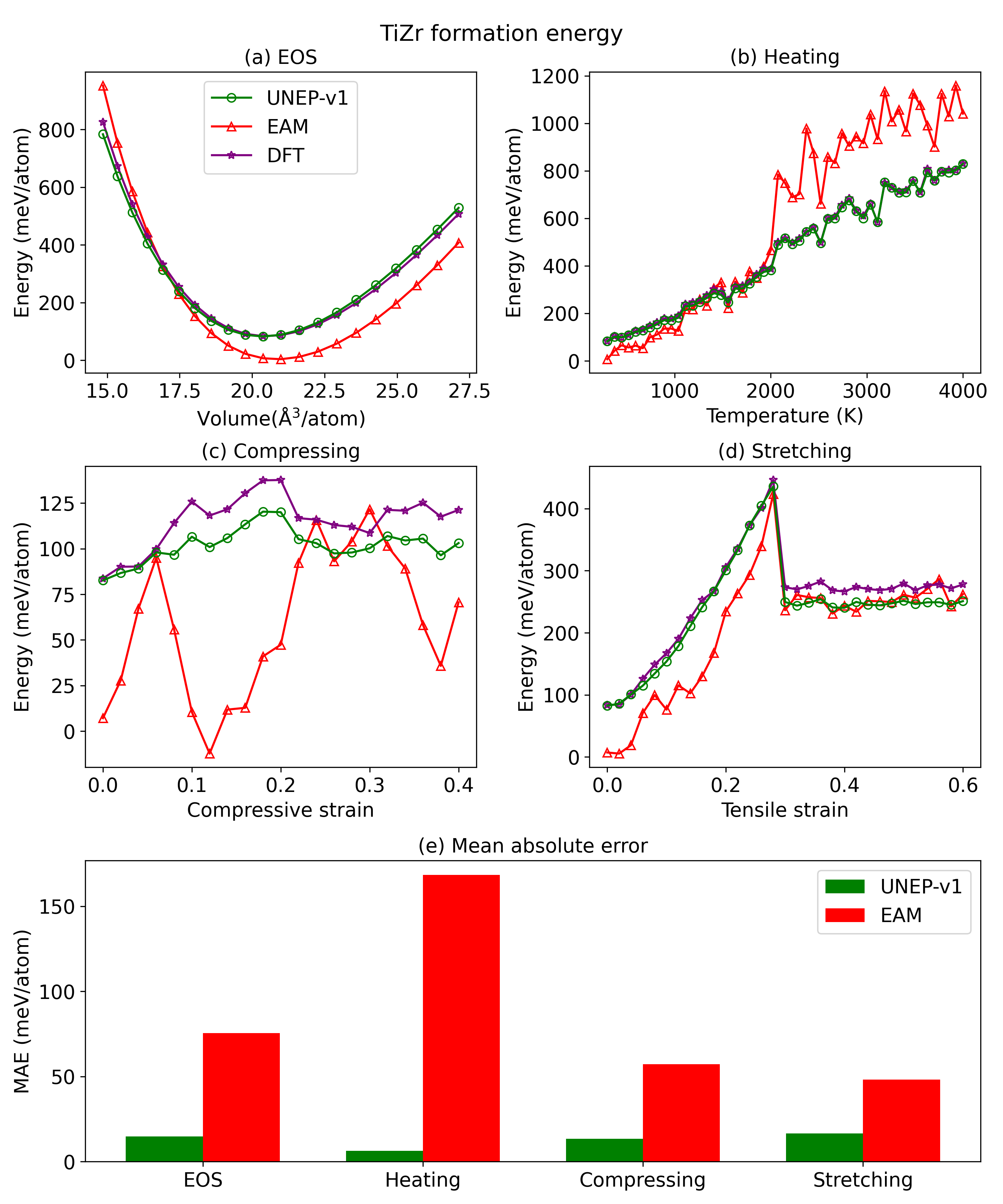}
\caption{
Similar to Fig.~\ref{fig:AlCu_results} but for HCP TiZr.
}
\label{fig:TiZr_results}
\end{figure}

\begin{figure}[b]
\centering
\includegraphics[width=\columnwidth]{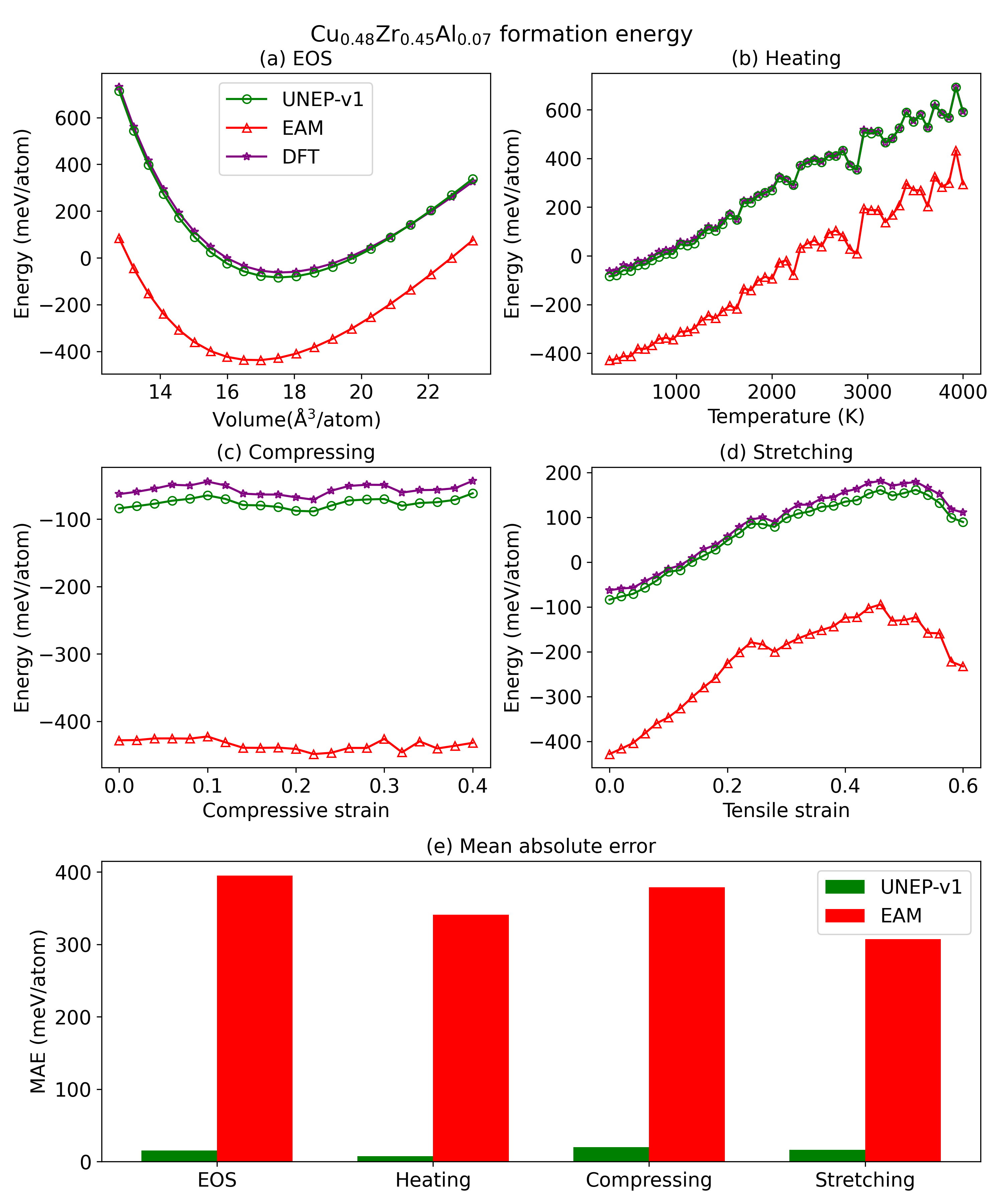}
\caption{
Similar to Fig.~\ref{fig:AlCu_results} but for metallic glassy Cu$_{0.475}$Zr$_{0.451}$Al$_{0.074}$.
}
\label{fig:Cu47.5Zr45.1Al7.4}
\end{figure}

\begin{figure}[b]
\centering
\includegraphics[width=\columnwidth]{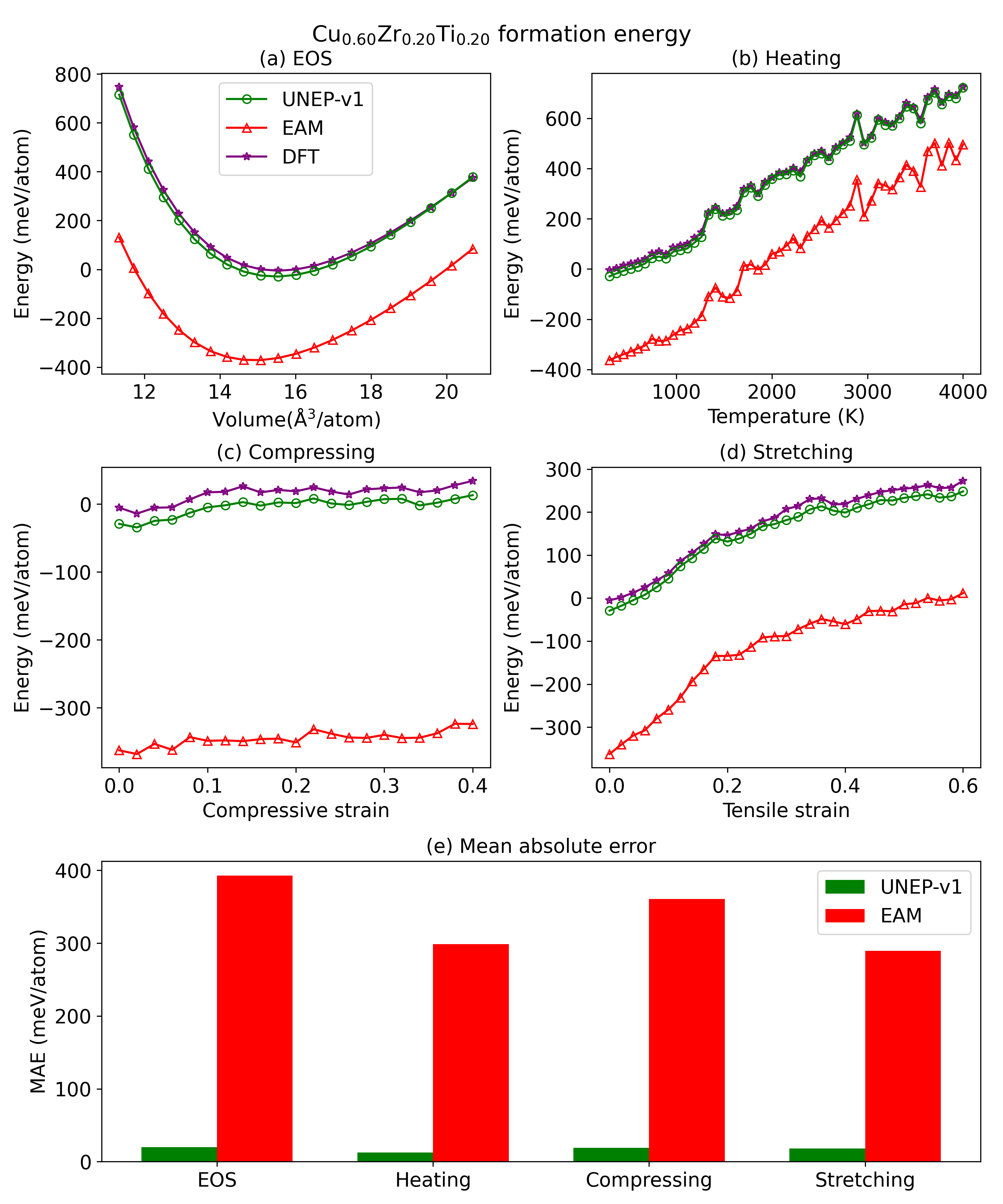}
\caption{
Similar to Fig.~\ref{fig:AlCu_results} but for metallic glassy Cu$_{0.6}$Zr$_{0.2}$Ti$_{0.2}$.
}
\label{fig:Cu60Zr20Ti20}
\end{figure}

\begin{figure}[b]
\centering
\includegraphics[width=\columnwidth]{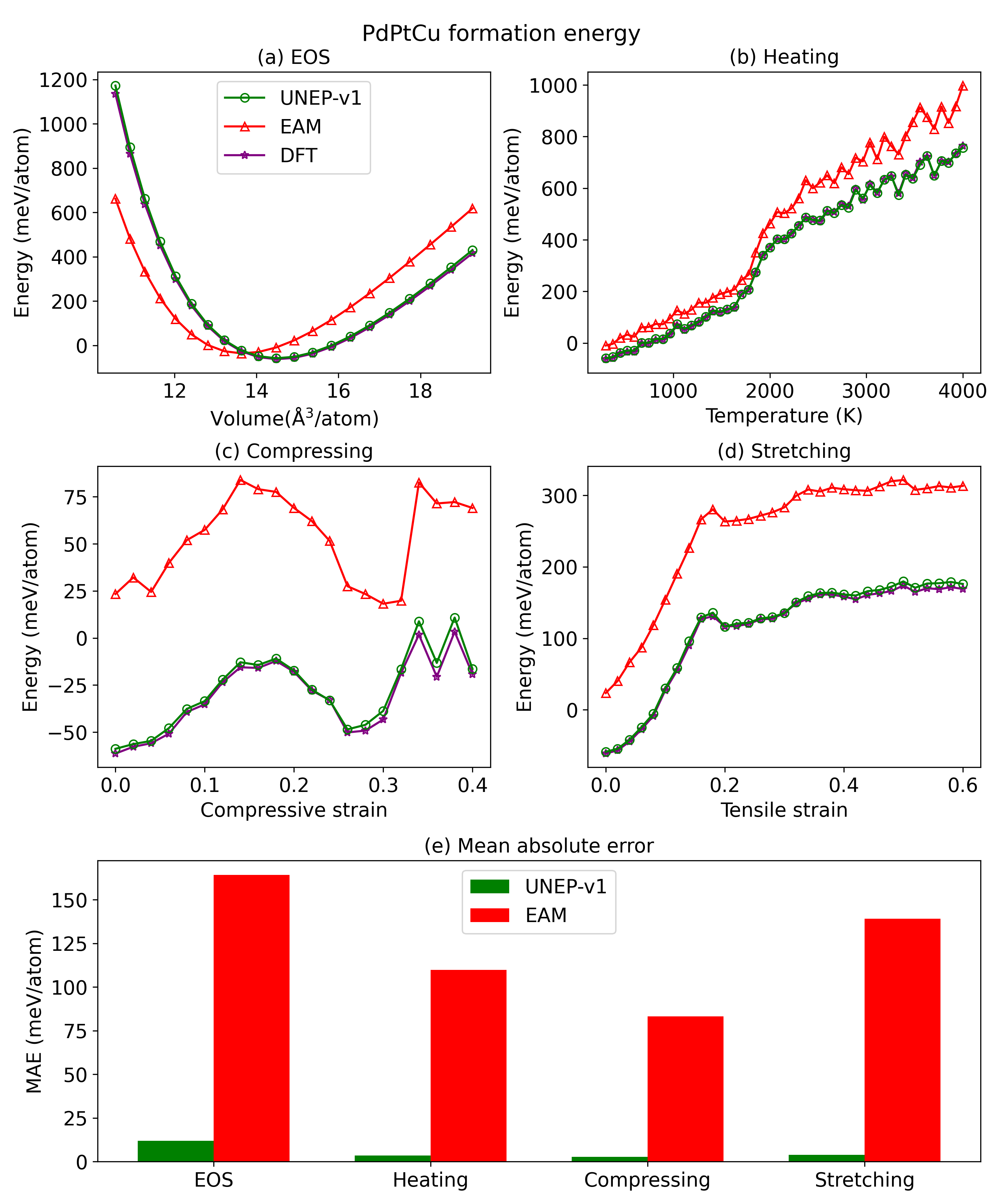}
\caption{
Similar to Fig.~\ref{fig:AlCu_results} but for FCC PdPtCu.
}
\label{fig:PdPtCu}
\end{figure}

\begin{figure}[b]
\centering
\includegraphics[width=\columnwidth]{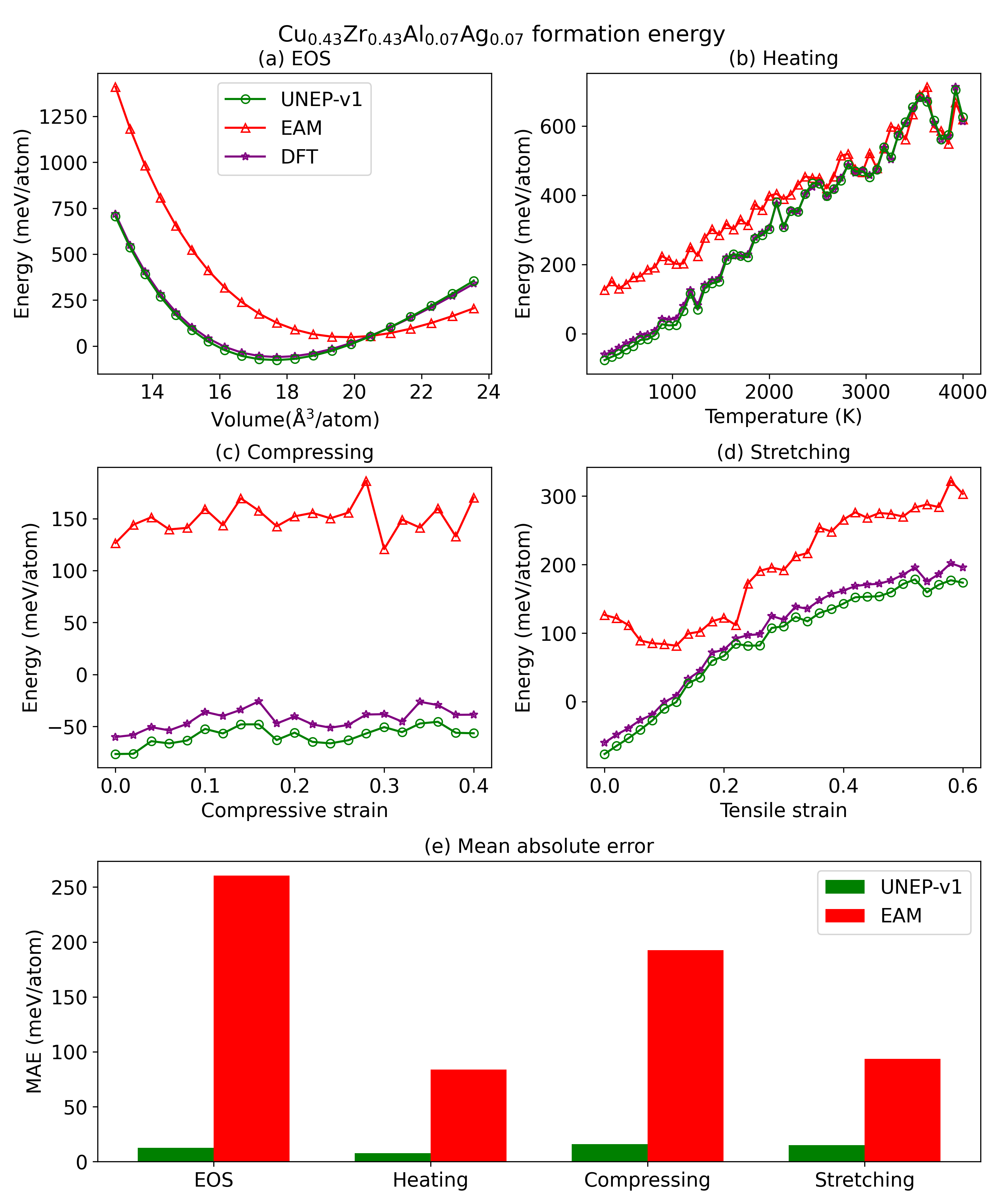}
\caption{
Similar to Fig.~\ref{fig:AlCu_results} but for metalltic glassy Cu$_{0.43}$Zr$_{0.43}$Al$_{0.07}$Ag$_{0.07}$.
}
\label{fig:Cu43Zr43Al7Ag7}
\end{figure}

\begin{figure}[b]
\centering
\includegraphics[width=\columnwidth]{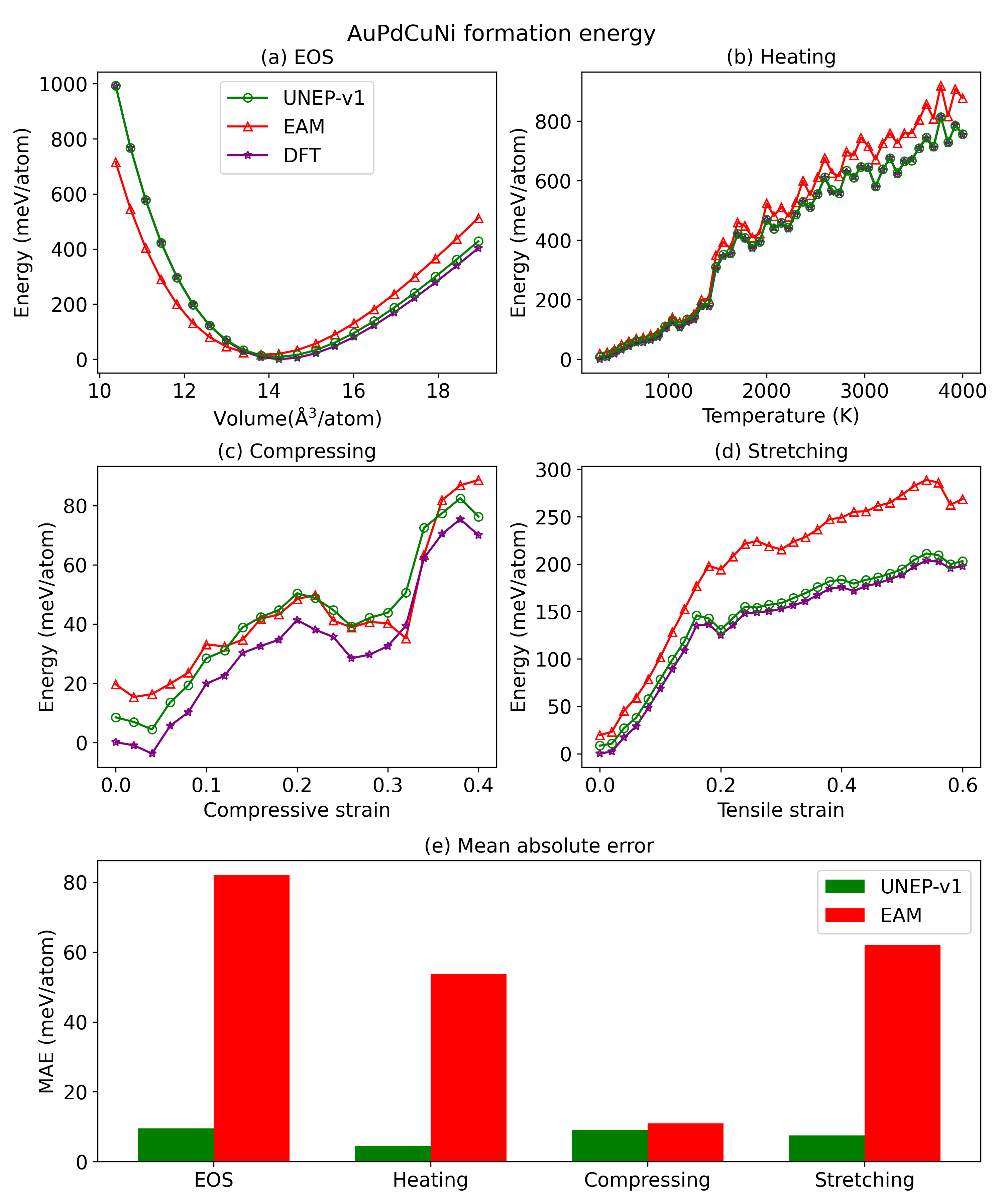}
\caption{
Similar to Fig.~\ref{fig:AlCu_results} but for FCC AuPdCuNi.
}
\label{fig:AuPdCuNi}
\end{figure}

\begin{figure}[b]
\centering
\includegraphics[width=\columnwidth]{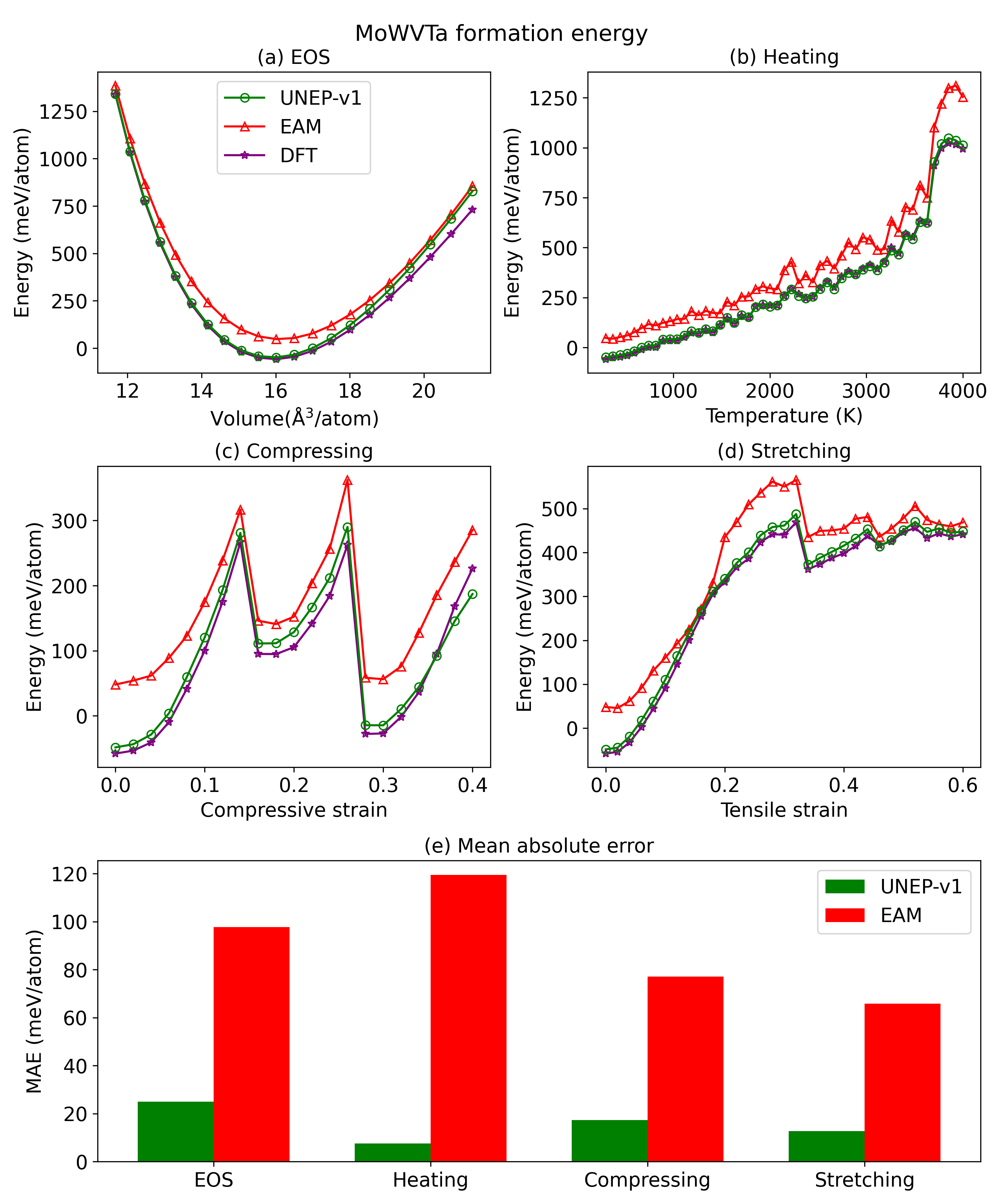}
\caption{
Similar to Fig.~\ref{fig:AlCu_results} but for BCC MoWVTa.
}
\label{fig:MoWVTa}
\end{figure}

\begin{figure}[b]
\centering
\includegraphics[width=\columnwidth]{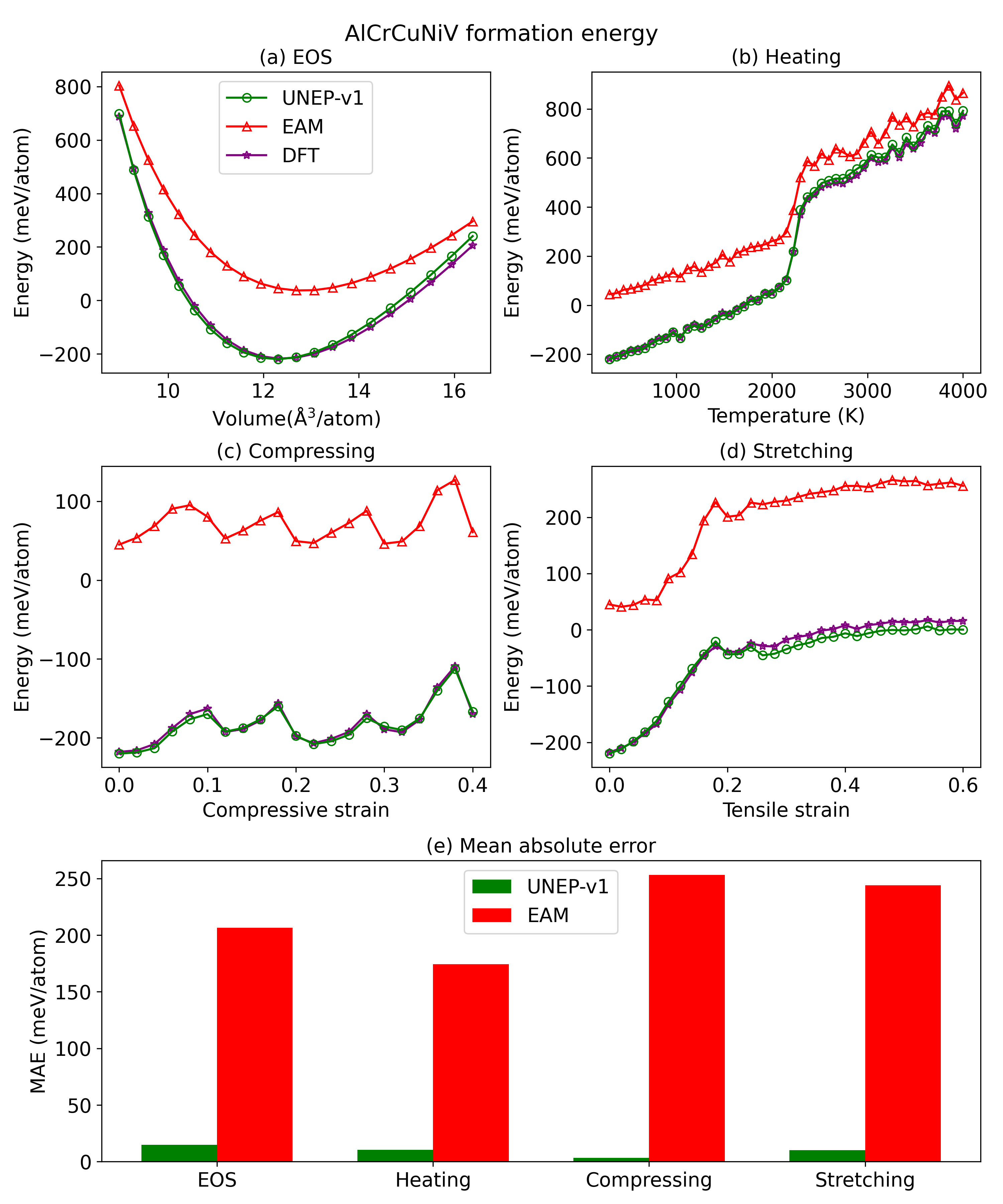}
\caption{
Similar to Fig.~\ref{fig:AlCu_results} but for BCC AlCrCuNiV.
}
\label{fig:AlCrCuNiV}
\end{figure}

\begin{figure}[b]
\centering
\includegraphics[width=\columnwidth]{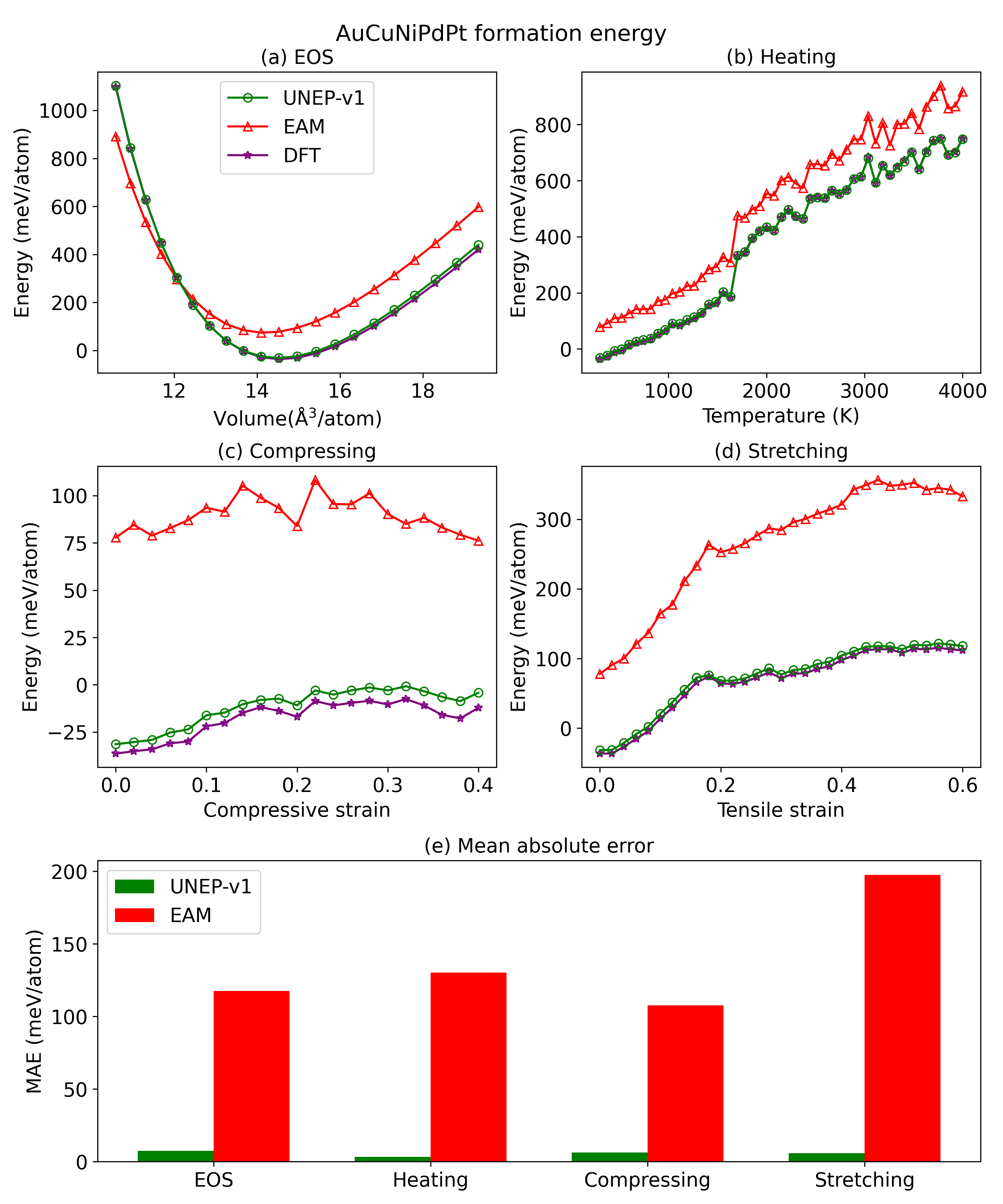}
\caption{
Similar to Fig.~\ref{fig:AlCu_results} but for FCC AuCuNiPdPt.
}
\label{fig:AuCuNiPdPt}
\end{figure}

\begin{figure}[b]
\centering
\includegraphics[width=\columnwidth]{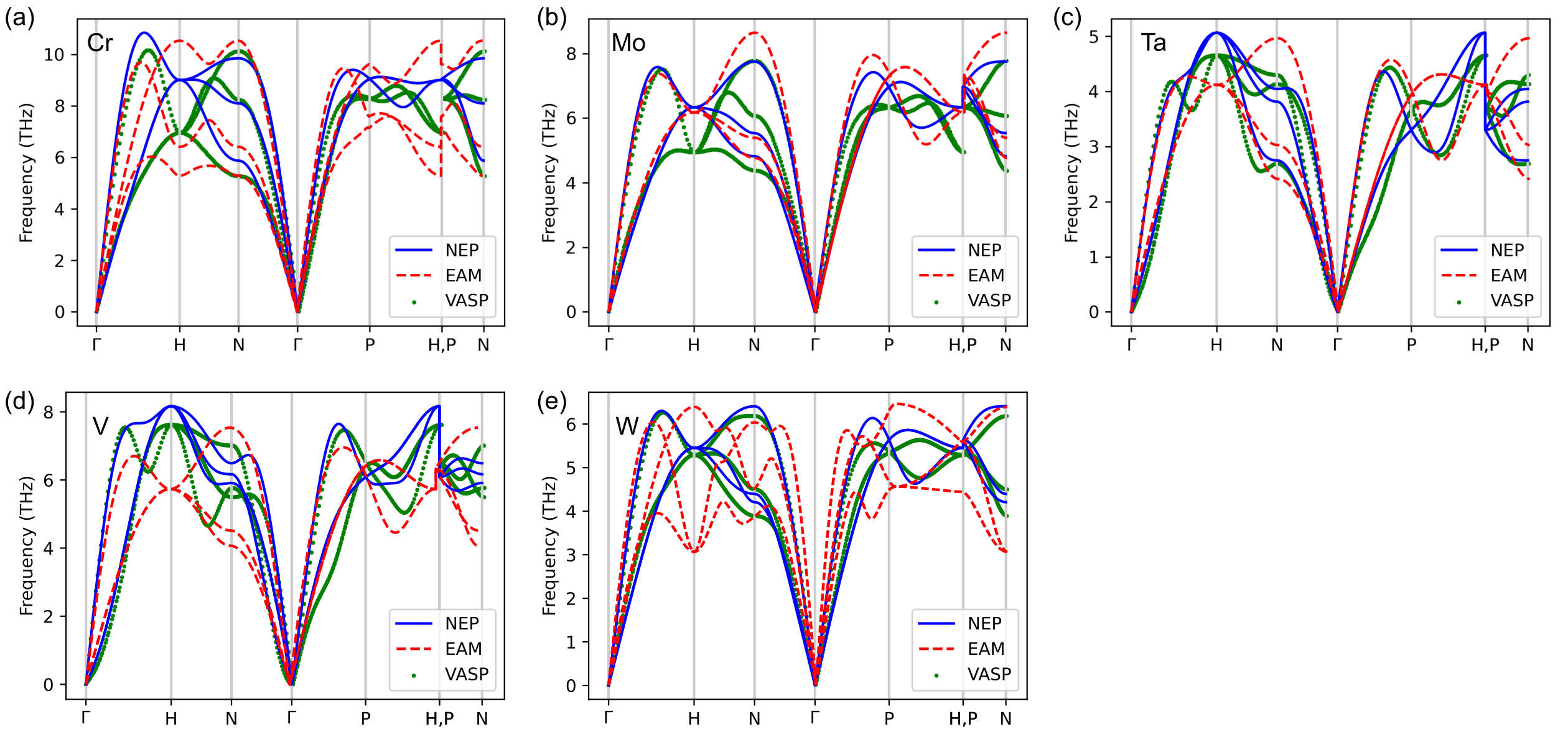}
\caption{
  Phonon dispersion relations for the BCC metals from UNEP-v1, EAM and DFT (VASP).
}
\label{fig:bcc_phonon}
\end{figure}

\begin{figure}
\centering
\includegraphics[width=\columnwidth]{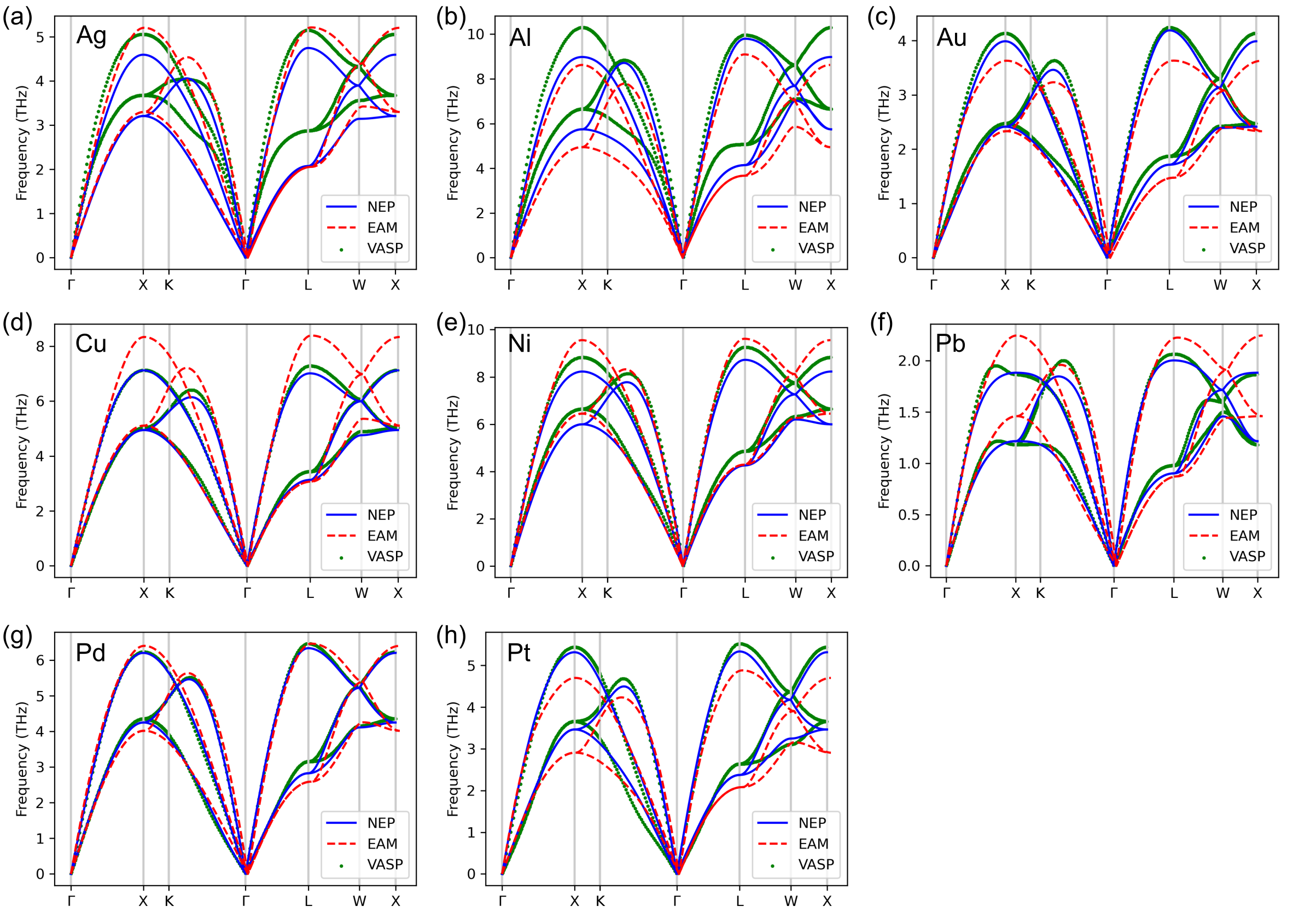}
\caption{
  Phonon dispersion relations for the FCC metals from UNEP-v1, EAM, and DFT (VASP).
}
\label{fig:fcc_phonon}
\end{figure}

\begin{figure}
\centering
\includegraphics[width=\columnwidth]{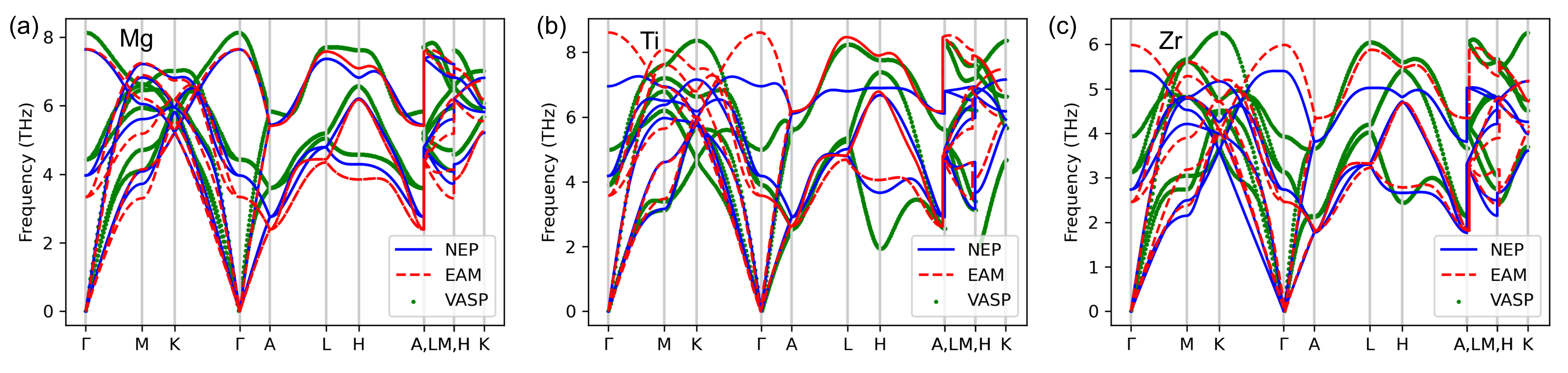}
\caption{
    Phonon dispersion relations for the HCP metals from UNEP-v1, EAM, and DFT (VASP).
}

\label{fig:hcp_phonon}
\end{figure}

\begin{figure}
\centering
\includegraphics[width=\columnwidth]{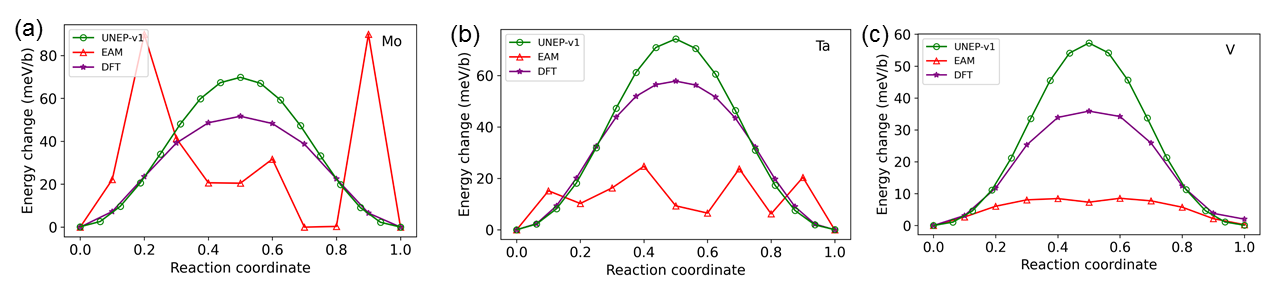}
\caption{
    Peierls barrier for the 1/2⟨111⟩ screw dislocation migration in elemental Mo, Ta, and V, predicted by UNEP-v1, EAM \cite{zhou2004prb}, and DFT.
}

\label{fig:screw}
\end{figure}

\begin{figure}[b]
\centering
\includegraphics[width=\columnwidth]{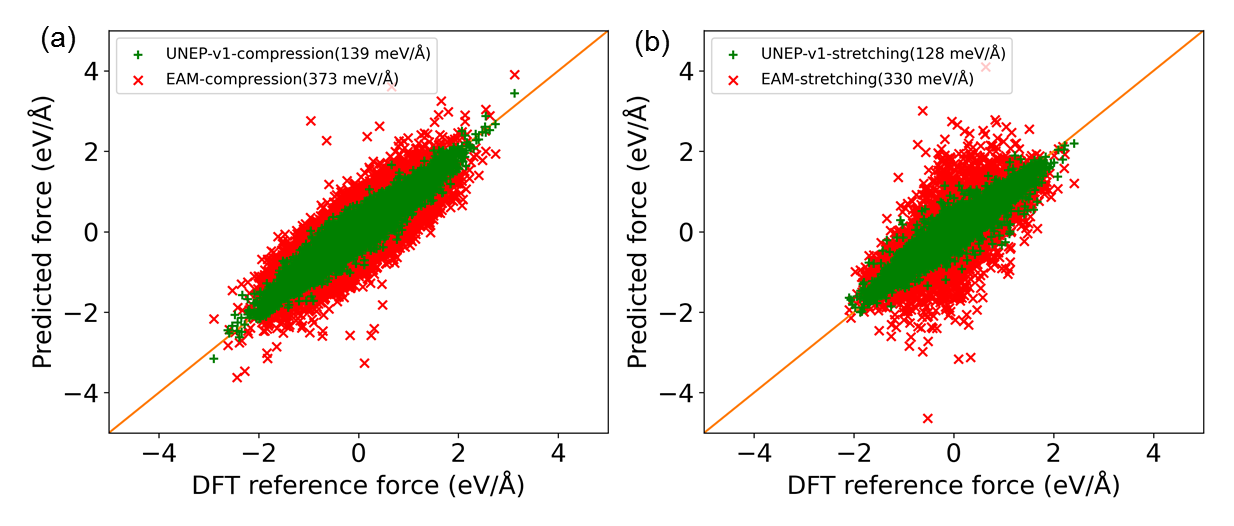}
\caption{
Parity plots for force predictions from UNEP-v1 and EAM \cite{zhou2004prb} compared to DFT for equimolar MoTaVW alloys sampled from various MD simulations using 256-atom supercells, including deformation processes up to 25\%\ compression (a) and stretching (b) at \qty{300}{\kelvin}. UNEP-v1 shows much better predictions than EAM \cite{zhou2004prb}, with much smaller force RMSEs as indicated in the legends.} 
\label{fig:MoVTaW-compress-stress}
\end{figure}

\begin{figure}[b]
\centering
\includegraphics[width=\columnwidth]{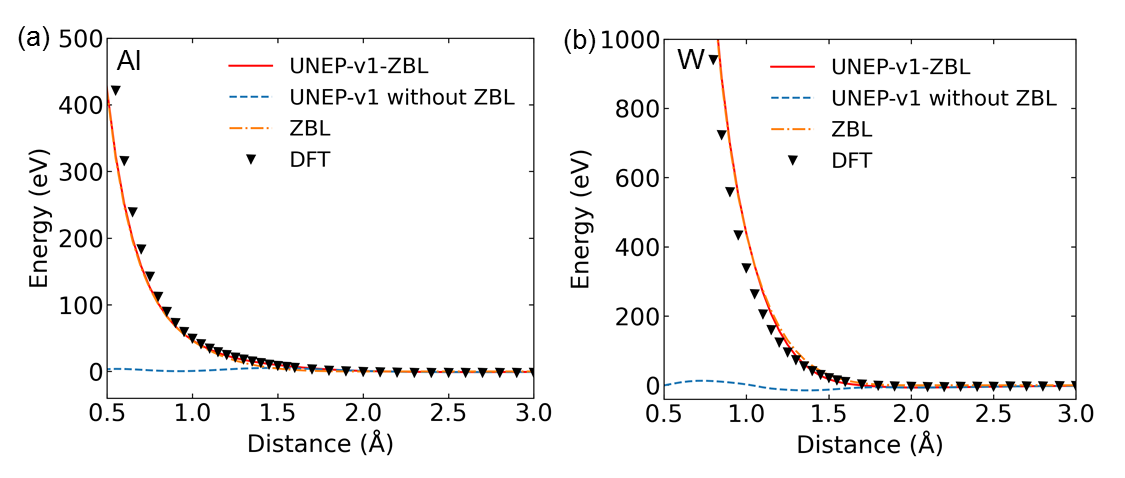}
\caption{Dimer energies as a function of atom distance for (a) Al and (b) W calculated using combined UNEP-v1 and ZBL (UNEP-v1-ZBL), UNEP-v1 without ZBL, ZBL, and DFT.
}
\label{fig:Dimer}
\end{figure}

\begin{figure}[htb]
\centering
\includegraphics[width=\columnwidth]{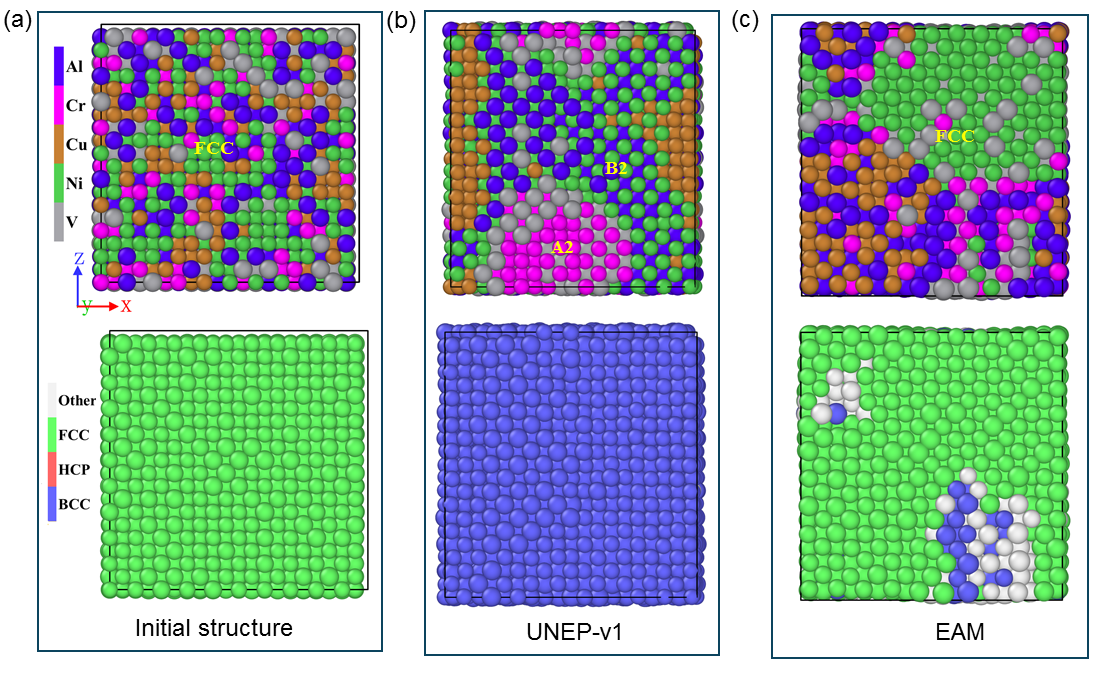}
\caption{
(a) Initial FCC structure of Al$_{0.20}$Cr$_{0.12}$Cu$_{0.19}$Ni$_{0.35}$V$_{0.14}$. (b)-(c) Snapshots of the final equilibrium structures from MCMD simulations using UNEP-v1 and EAM \cite{zhou2004prb}. Upper panels show the atomistic structures, while lower panels display the corresponding common neighbor analysis. UNEP-v1 successfully produces both disordered (A2) and ordered (B2) BCC structures in full agreement with experiments \cite{YI2020jac}. In contrast, EAM potential by Zhou \textit{et al.} \cite{zhou2004prb} keeps the system mostly in FCC structure, unable to reproduce the experimentally expected BCC structure.
}
\label{fig:AlCrCuNiV2}
\end{figure}

\begin{figure}
\centering
\includegraphics[width=\columnwidth]{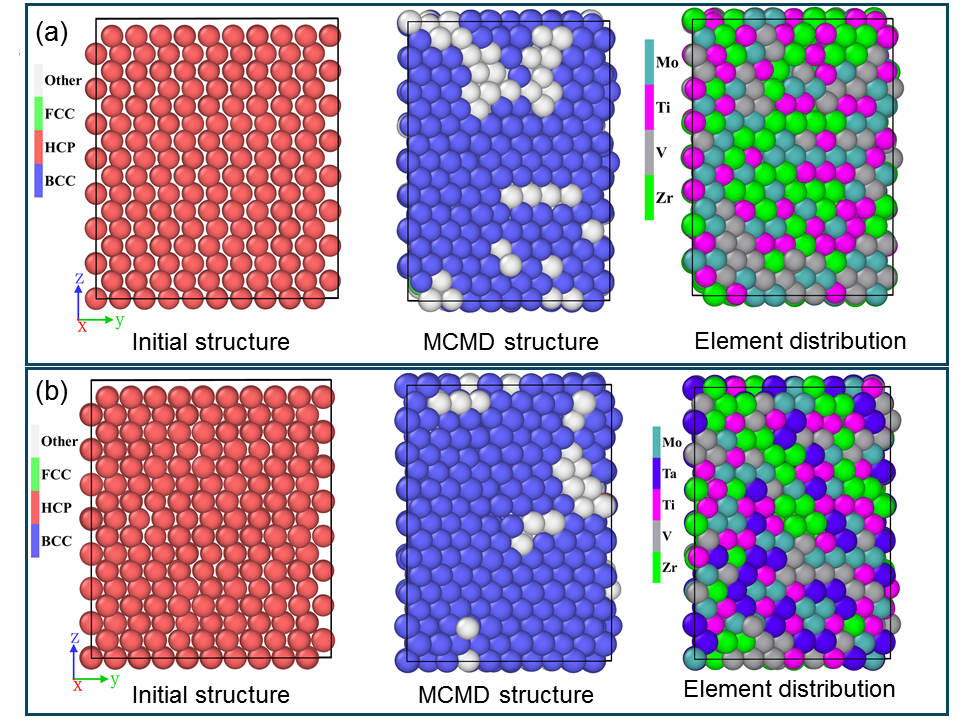}
\caption{Upper panels demonstrate the transformation of the equimolar TiZrVMo alloy from the initial HCP structure to the BCC structure during MCMD simulations using UNEP-v1 model. Lower panels show a similar transformation for equimolar TiZrVMoTa alloy from the initial HCP structure to the BCC structure during MCMD simulations using UNEP-v1 model. These transformations and the stable phases are in accordance with experimental observations~\cite{Mu2017@jac}, suggesting that sampling binary alloys can correctly capture phase transitions occurring in multi-component alloys.}
\label{fig:hcp}
\end{figure}

\begin{figure}
\centering
\includegraphics[width=\columnwidth]{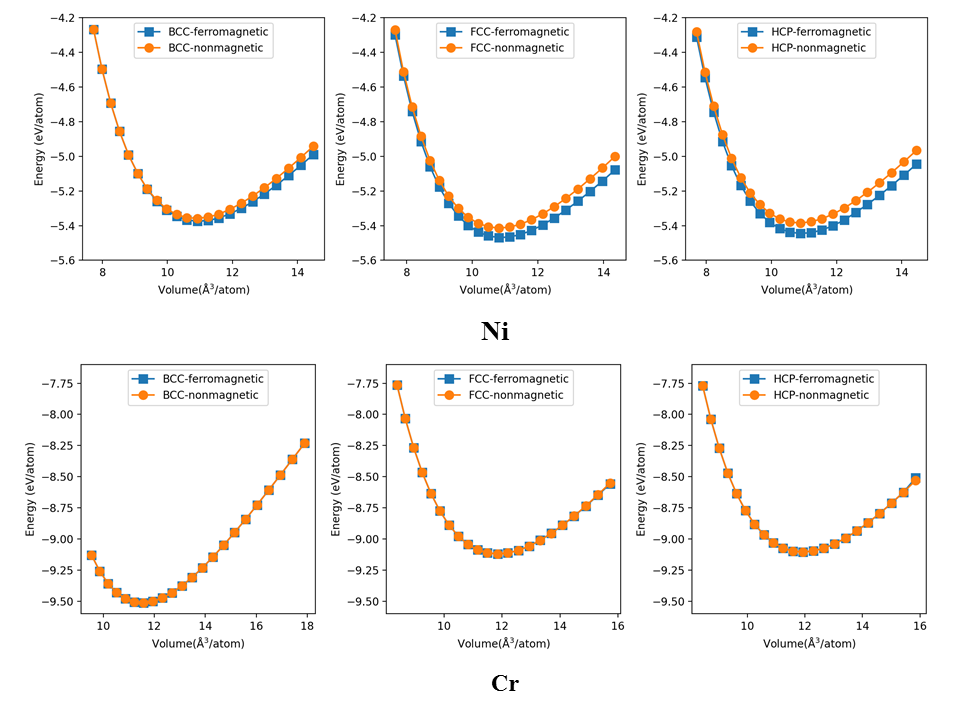}
\caption{Comparison of the energies for Ni and Cr in BCC, FCC, and HCP structures from DFT calculations with and without considering magnetism.}
\label{fig:spin}
\end{figure}


\clearpage
\section{\listtablename{}}
\newcommand{\tabcite}[1]{\ensuremath{^\text{[\!\!\!\citenum{#1}]}}}

\begin{table}[thb]
\centering
\setlength{\tabcolsep}{1.8Mm}
\caption{
    Elastic constant (in units of GPa).
}
\label{table:elastic_constant}
\begin{tabular}{ccrrr|ccrrr}
\hline
\hline
Species             & Component  & DFT~\cite{De2015} & EAM   & UNEP-v1    & Element             & Component  & DFT~\cite{De2015} & EAM   & UNEP-v1   \\
\hline
\multirow{3}{*}{Ag} & C11        & 100    & 124.7 & 109.6  & \multirow{3}{*}{Al}  & C11        & 104    & 106.5 & 121.3 \\
                    & C12        & 82     & 93.4  & 81.7   &                      & C12        & 73     & 59.5  & 54  \\
                    & C44        & 41     & 46.1  & 41.9   &                      & C44        & 32     & 27.9  & 39  \\
\multirow{3}{*}{Au} & C11        & 144    & 186.7 & 154.8  &  \multirow{3}{*}{Cu} & C11        & 180    & 170.0   & 175.4 \\
                    & C12        & 134    & 156.9 & 118.5  &                      & C12        & 127    & 121.0   & 128.3 \\
                    & C44        & 29     & 42.1  & 32.7   &                      & C44        & 78     & 75.5  & 78.8  \\
\multirow{3}{*}{Ni} & C11        & 276    & 245.6 & 273  & \multirow{3}{*}{Pb}  & C11        & 47     & 49.6  & 49  \\
                    & C12        & 159    & 147.0  & 179.2  &                     & C12        & 32     & 42.1  & 36.4  \\
                    & C44        & 132    & 124.2 & 112.3  &                      & C44        & 18     & 15    & 15  \\
\multirow{3}{*}{Pd} & C11        & 187    & 233.9 & 190.6  &  \multirow{3}{*}{Pt} & C11        & 303    & 346.4 & 305.2 \\
                    & C12        & 147    & 175.0   & 149.7  &                    & C12        & 220    & 249.1 & 227.2 \\
                    & C44        & 71     & 71.1  & 71.8  &                      & C44        & 54     & 75.8  & 71.2  \\
\multirow{3}{*}{Cr} & C11        & 499    & 304.6 & 577.4  & \multirow{3}{*}{Mo}  & C11        & 472    & 455.2 & 460.9\\
                    & C12        & 139    & 135.8 & 122.9    &                    & C12        & 158    & 167.3 & 174.2   \\
                    & C44        & 102    & 122.8 & 84.4   &                      & C44        & 106    & 113.2 & 88  \\
\multirow{3}{*}{Ta} & C11        & 265    & 262.0   & 261.1  & \multirow{3}{*}{V} & C11        & 276    & 231.6 & 333.5    \\
                    & C12        & 158    & 158.2  & 165.6  &                     & C12        & 131    & 119.8 & 123.4 \\
                    & C44        & 69     & 82.4  & 47    &                      & C44        & 16     & 45.9  & 18.4     \\
\multirow{3}{*}{W}  & C11        & 510    & 520.9 & 505   &   \multirow{3}{*}{} &            &     &   & \\
                    & C12        & 201    & 205.0   & 208.2   &                   &            &     &   &  \\
                    & C44        & 143    & 160.9 & 121.5 &                       &            &     &   &  \\
\multirow{6}{*}{Mg} & C11        & 58     & 54.1  & 63   & \multirow{6}{*}{Ti} & C11         & 177    & 135.3 & 173 \\
                    & C12        & 30     & 30.0    & 28.3   &                   & C12         & 83     & 94.5  & 93.7    \\
                    & C13        & 22     & 20.5  & 20.3  &                      & C13         & 76     & 68.5  & 68.6  \\
                    & C33        & 66     & 67.7  & 72.5  &                      & C33         & 191    & 203   & 207.3  \\
                    & C44        & 20     & 13.3  & 17  &                      & C44         & 42     & 30.3  & 40.6  \\
                    & C66        & 14     & 12.0    & 17.3  &                    & C66         & 47     & 20.4 & 39.6 \\
\multirow{6}{*}{Zr} & C11        & 144    & 118.3 & 141.4 &   \multirow{6}{*}{}  &       &    &  &  \\
                     & C12       & 65     & 88.9  & 81.4    &                     &       &     &  &  \\
                     & C13       & 67     & 67.0    & 62.2  &                      &        &     &    &  \\
                     & C33       & 162    & 181.1 & 137.4  &                     &        &   &  &  \\
                     & C44       & 26     & 24.5  & 21.3&                        &      &      &   &  \\
                     & C66       & 40     & 14.7  & 30&                          &       &       &   &   \\
\hline
\hline
\end{tabular}
\end{table}

\begin{table}[thb]
\centering
\setlength{\tabcolsep}{6Mm}
\caption{
    Mono-vacancy formation energy (in units of eV).
}
\label{table:vacancy_energy}
\begin{tabular}{cccccc}
\hline
\hline
Species & DFT & EAM & UNEP-v1  \\
\hline
Ag	&	0.96	&	1.10    &	0.80		\\
Al	&	0.55	&	0.65	&	0.61		\\
Au	&	0.51	&	0.99	&	0.57		\\
Cu	&	1.12	&	1.28	&	1.03		\\
Ni	&	1.41	&	1.70    &	1.42		\\
Pb	&	0.39	&	0.59	&	0.44		\\
Pd	&	1.23	&	1.55	&	1.20		\\
Pt	&	0.68	&	1.54	&	1.03    	\\
Cr	&	2.54	&	2.06	&	3.04		\\
Mo	&	2.77	&	2.95	&	3.07		\\
Ta	&	2.82	&	2.97	&	2.79		\\
V	&	2.28	&	2.14	&	2.35		\\
W	&	3.34	&	3.58	&	3.73		\\
Mg	&	0.85	&	0.65	&	0.80		\\
Ti	&	2.00	&	1.63	&	2.40		\\
Zr	&	2.00	&	1.84	&	2.20		\\
\hline
\hline
\end{tabular}
\end{table}

\begin{table}[thb]
\centering
\caption{
    Surface formation energy (in units of J/m$^2$).
}
\label{table:surface_energy}
\begin{tabular*}{\textwidth}{@{\extracolsep{\fill}}cccccccccccc}
\hline
\hline
\multicolumn{1}{l}{Species}        & \multicolumn{3}{c}{DFT~\cite{Tran2016surface}} & \multicolumn{3}{c}{EAM} & \multicolumn{3}{c}{UNEP-v1} \\
\cline{2-4} \cline{5-7} \cline{8-10}
\multicolumn{1}{l}{Surface} & 100    & 110    & 111   & 100    & 110    & 111   & 100    & 110    & 111   \\
\hline
Ag & 0.82 & 0.87 & 0.76 & 0.99 & 1.11 & 0.91 & 0.83 & 0.89 & 0.73\\
Al & 0.91 & 0.98 & 0.77 & 0.93 & 1.04 & 0.91 & 0.93 & 0.98 & 0.86 \\
Au & 0.86 & 0.91 & 0.71 & 1.01 & 1.11 & 0.90 & 0.88 & 0.89 & 0.72 \\
Cu & 1.47 & 1.56 & 1.34 & 1.58 & 1.77 & 1.51 & 1.49 & 1.58 & 1.32 \\
Ni & 2.21 & 2.29 & 1.92 & 1.90 & 2.08 & 1.79 & 2.23 & 2.34 & 1.98 \\
Pb & 0.33 & 0.33 & 0.26 & 0.39 & 0.43 & 0.35 & 0.29 & 0.32 & 0.26 \\
Pd & 1.52 & 1.57 & 1.36 & 1.64 & 1.81 & 1.52 & 1.50 & 1.59 & 1.34 \\
Pt & 1.86 & 1.87 & 1.49 & 2.19 & 2.51 & 2.08 & 1.77 & 1.78 & 1.43 \\
Cr & 3.63 & 3.22 & 3.44 & 1.86 & 1.69 & 2.11 & 3.63 & 3.11 & 3.58 \\
Mo & 3.18 & 2.78 & 2.96 & 2.48 & 2.16 & 2.77 & 3.13 & 2.70 & 3.13 \\
Ta & 2.47 & 2.34 & 2.70 & 2.35 & 2.00 & 2.60 & 2.77 & 2.40 & 2.84 \\
V  & 2.38 & 2.41 & 2.70 & 1.94 & 1.66 & 2.15 & 2.78 & 2.46 & 2.79 \\
W  & 3.95 & 3.23 & 3.47 & 2.99 & 2.57 & 1.83 & 3.89 & 3.19 & 3.72 \\
\hline
{Surface} & 0001    & 1010    & 1011   & 0001    & 1010    & 1011   & 0001    & 1010    & 1011   \\
\hline
Mg & 0.51 & 0.60 & 0.63 & 0.38 & 0.40 & 0.40 & 0.61 & 0.64 & 0.65 \\
Ti & 2.15 & 2.22 & 2.25 & 1.83 & 1.89 & 1.64 & 2.30 & 2.26 & 2.37 \\
Zr & 1.60 & 1.66 & 1.57 & 1.26 & 1.35 & 1.34 & 1.42 & 1.52 & 1.50 \\
\hline
\hline
\end{tabular*}
\end{table}

\begin{table}[thb]
\centering
\setlength{\tabcolsep}{6Mm}
\caption{
    Melting temperature (in units of K).
}
\label{table:melting_temperature}
\begin{tabular}{crrrrr}
\hline
\hline
Species & EAM & UNEP-v1 & Exp. \cite{CRChandbook} \\
\hline
Ag & 1135 & 970  & 1235 \\
Al & 553  & 840  & 933  \\
Au & 1115 & 900  & 1337 \\
Cu & 1161 & 1220 & 1358 \\
Ni & 1500 & 1670 & 1728 \\
Pb & 611  & 550  & 601  \\
Pd & 1553 & 1510 & 1828 \\
Pt & 1422 & 1610 & 2041 \\
Cr & 2128 & 2520 & 2180 \\
Mo & 3370 & 2870 & 2896 \\
Ta & 2950 & 3150 & 3290 \\
V  & 1664 & 2240 & 2183 \\
W  & 4221 & 3710 & 3695 \\
Mg & 682  & 860  & 923  \\
Ti & 1578 & 1610 & 1941 \\
Zr & 1752 & 1860 & 2128 \\
\hline
\hline
\end{tabular}
\end{table}

\clearpage

\newpage

\phantomsection


\end{document}